\documentclass[traditabstract]{aa}

\usepackage{amsmath}
\usepackage{natbib}
\usepackage{graphicx}
\usepackage{txfonts}
\usepackage{xspace}

\bibpunct{(}{)}{;}{a}{}{,}

\newcommand{\gro}{GRO~J1008$-$57\xspace}

\newcommand{\RXTE}{\textsl{RXTE}\xspace}
\newcommand{\Swift}{\textsl{Swift}\xspace}
\newcommand{\Suzaku}{\textsl{Suzaku}\xspace}
\newcommand{\CGRO}{\textsl{CGRO}\xspace}
\newcommand{\INTEGRAL}{\textsl{INTEGRAL}\xspace}

\graphicspath{{./plots/}}

\begin{document}

\title{GRO~J1008$-$57: an (almost) predictable transient X-ray binary}

\author{
      Matthias K\"uhnel \inst{1}
 \and Sebastian M\"uller \inst{1}
 \and Ingo Kreykenbohm \inst{1}
 \and Felix F\"urst \inst{2}
 \and Katja Pottschmidt \inst{3,4}
 \and Richard E. Rothschild \inst{5}
 \and Isabel Caballero \inst{6}
 \and Victoria Grinberg \inst{1}
 \and Gabriele Sch\"onherr \inst{7}
 \and Chris Shrader\inst{4,8}
 \and Dmitry Klochkov \inst{9}
 \and R\"udiger Staubert \inst{9}
 \and Carlo Ferrigno \inst{10}
 \and \mbox{Jos\'e-Miguel Torrej\'on} \inst{11}
 \and Silvia Mart\'inez-N\'u\~nez\inst{11}
 \and J\"orn Wilms \inst{1}
}

\authorrunning{K\"uhnel et al.}

\institute{
      Dr. Karl Remeis-Observatory \& ECAP, Universit\"at
      Erlangen-N\"urnberg, Sternwartstr. 7, 96049 Bamberg, Germany 
 \and Space Radiation Lab, California Institute of Technology, MC
      290-17 Cahill, 1200 E. California Blvd., Pasadena, CA 91125, USA 
 \and CRESST, Center for Space Science and Technology, UMBC,
      Baltimore, MD 21250, USA
 \and NASA Goddard Space Flight Center, Greenbelt, MD 20771, USA
 \and Center for Astronomy and Space Sciences, University of
      California, San Diego, La Jolla, CA 92093, USA 
 \and CEA Saclay, DSM/IRFU/SAp-UMR AIM (7158) CNRS/CEA/Universit\'e
      Paris 7, Diderot, 91191 Gif sur Yvette, France 
 \and Leibniz-Institut f\"ur Astrophysik Potsdam, An der Sternwarte
      16, 14482 Potsdam, Germany 
 \and Universities Space Research Association, Columbia, MD 21044, USA
 \and Institut f\"ur Astronomie und Astrophysik, Universit\"at
      T\"ubingen, Sand 1, 72076 T\"ubingen, Germany
 \and ISDC Data Center for Astrophysics, Chemin d'\'Ecogia 16, 1290
      Versoix, Switzerland
 \and Instituto de F\'isica Aplicada a las Ciencias y las
      Tecnolog\'ias, Universidad de Alicante, 03080 Alicante, Spain   
}

\abstract{A study of archival \RXTE, \Swift, and \Suzaku pointed
  observations of the transient high mass X-ray binary \gro is presented.
  A new orbital ephemeris based on pulse arrival timing shows times of
  maximum luminosities during outbursts of \gro to be close to
  periastron at orbital phase $-0.03$. This makes the source one of a
  few for which outburst dates can be predicted with very high
  precision.

  Spectra of the source in 2005, 2007, and 2011 can be well described
  by a simple power law with high energy cutoff and an additional black
  body at lower energies. The photon index of the power law and the
  black body flux only depend on the 15--50\,keV source flux. No
  apparent hysteresis effects are seen. These correlations allow to
  predict the evolution of the pulsar's X-ray spectral shape over all
  outbursts as a function of just one parameter, the source's flux.
  If modified by an additional soft component, this prediction even
  holds during \gro's 2012 type~II outburst.}

\date{Received: --- / Accepted: ---}

\maketitle

\section{Introduction} \label{sec:intro}

\object{GRO~J1008$-$57} was discovered by \CGRO during an X-ray
outburst on 1993 July 14 \citep{wilson1994a,stollberg1993a}. It is a
transient neutron star with a Be-star companion of type B0e
\citep{coe1994a} at a distance of 5.8\,kpc \citep{riquelme2012a}. In
such high-mass X-ray binaries (HMXBs) the compact object is on a wide
eccentric orbit around its companion, which itself features a
circumstellar disc due to its fast rotation. Periastron passages of
the neutron star lead to accretion from the donor's decretion disk and
regular X-ray outbursts. Based on \RXTE-ASM measurements of the X-ray
light curve, \citet[][see also \citealt{levine2011a}]{levine2006a}
derived an orbital period of 248.9(5)\,d\footnote{The error bars are
  given in units of the last digit shown.}. Since individual X-ray
outbursts are relatively short, with durations of around 14\,d,
determining the parameters of the binary orbit is challenging.
\citet{coe2007a} have derived an orbital solution (see
Table~\ref{tab:orbit}) based on pulse period folding of the 93.6\,s
X-ray period \citep{stollberg1993a}.

The hard X-ray spectrum of \gro was first studied by
\citet{grove1995a} and \citet{shrader1999a}. Above 20\,keV the
spectrum can be well described by an exponentially cutoff power law
\citep{grove1995a,shrader1999a,coe2007a}. Below 20\,keV, \Suzaku
observations show a complex spectrum with Fe line fluorescent emission
and a power law continuum \citep{naik2011a}.

A slight deviation from the continuum at 88\,keV in \CGRO-OSSE
observations was interpreted by \citet{shrader1999a} as a possible
cyclotron line feature, i.e., as a feature caused by inelastic
scattering of photons off electrons quantized in the strong magnetic
field at the neutron star's poles \citep[][and references
therein]{schoenherr2007a,caballero2012b,pottschmidt2012a}. Even if
the 88\,keV feature is the second harmonic cyclotron line as argued by \citet{shrader1999a},
\gro would be one of the most strongly magnetized neutron star in an accreting system to date.

In this paper an analysis of archival observations of
\gro with \RXTE, \Suzaku, and \Swift is presented.
Section~\ref{sec:data} presents the data analysis strategy. In
Sect.~\ref{sec:timing} an updated orbital solution based on data from
outbursts in 2005 and 2007 is presented. The X-ray spectrum of \gro is
discussed in Sect.~\ref{sec:spectral} and the spectroscopic results
are applied and compared in Sect.~\ref{sec:dec2011}. The paper
closes with a summary and conclusions in Sect.~\ref{sec:conclude}.

\section{Observations \& Data Reduction} \label{sec:data}

\begin{table}
 \centering
 \caption{Log of observations. The table includes the satellite used
   (Sat.), the observation ID, the Modified Julian Date of the start
   of observation, and the Exposure (Exp.) in seconds. The last column
   (E) marks data epochs, for which spectra are combined for spectral
   analysis. Horizontal lines separate different outbursts or campaigns.} 
 \label{tab:obsids}
 \begin{tabular}{ccccc}
  Sat. & Observation & Starttime & Exp. & E \\ \hline\hline
  \multicolumn{5}{@{}l}{Observations in quiescence (1996/1997)}\\
  RXTE & 20132-01-01-000 & 50412.49 & $\phantom{0}$9439 & I \\
RXTE & 20132-01-01-00 & 50412.80 & $\phantom{0}$6623 & I \\
RXTE & 20132-01-01-01 & 50413.05 & $\phantom{0}$8719 & I \\
RXTE & 20132-01-01-02 & 50413.21 & $\phantom{0}$1136 & I \\
RXTE & 20132-01-01-04 & 50413.69 & $\phantom{0}$4319 & I \\
RXTE & 20132-01-01-05 & 50413.84 & $\phantom{0}$1472 & I \\

  RXTE & 20132-01-02-00 & 50466.74 & 17135 & I \\
RXTE & 20132-01-02-01 & 50467.09 & $\phantom{0}$4687 & I \\
RXTE & 20132-01-02-02 & 50467.22 & $\phantom{0}$5023 & I \\
RXTE & 20132-01-02-03 & 50467.38 & $\phantom{0}$3648 & I \\
RXTE & 20132-01-02-04 & 50467.64 & $\phantom{0}$7327 & I \\
RXTE & 20132-01-03-000 & 50514.36 & 13327 & I \\
RXTE & 20132-01-03-00 & 50514.67 & $\phantom{0}$9999 & I \\
RXTE & 20132-01-03-01 & 50514.87 & 10047 & I \\
RXTE & 20132-01-03-02 & 50515.02 & 11231 & I \\
RXTE & 20123-09-01-00 & 50519.20 & $\phantom{0}$4335 & I \\
RXTE & 20123-09-02-00 & 50537.48 & $\phantom{0}$4559 & I \\
RXTE & 20123-09-03-00 & 50565.90 & $\phantom{0}$5375 & I \\
RXTE & 20132-01-04-00 & 50567.38 & 15615 & I \\
RXTE & 20132-01-04-03 & 50568.36 & $\phantom{0}$9455 & I \\
RXTE & 20132-01-04-01 & 50569.52 & $\phantom{0}$7855 & I \\
RXTE & 20132-01-04-02 & 50569.68 & $\phantom{0}$9599 & I \\
RXTE & 20132-01-04-04 & 50569.95 & $\phantom{0}$1456 & I \\
RXTE & 20123-09-04-00 & 50602.11 & $\phantom{0}$4607 & I \\
RXTE & 20123-09-05-00 & 50619.11 & $\phantom{0}$4303 & I \\

  \multicolumn{5}{@{}l}{Outburst in 2005 February}\\
  RXTE & 90089-03-01-00 & 53421.56 & $\phantom{00}$832 \\
RXTE & 90089-03-02-01 & 53426.01 & $\phantom{0}$2208 & II \\
RXTE & 90089-03-02-08 & 53426.29 & $\phantom{00}$863 & II \\
RXTE & 90089-03-02-07 & 53426.40 & $\phantom{0}$1232 & II \\
RXTE & 90089-03-02-02 & 53427.11 & $\phantom{0}$2912 & III \\
RXTE & 90089-03-02-000 & 53427.29 & 15536 & III \\
RXTE & 90089-03-02-00 & 53427.59 & $\phantom{0}$7792 & III \\
RXTE & 90089-03-02-03 & 53428.11 & $\phantom{0}$1808 & IV \\
RXTE & 90089-03-02-04 & 53428.27 & $\phantom{0}$5968 & IV\\
RXTE & 90089-03-02-05 & 53428.84 & $\phantom{00}$735 & IV \\
RXTE & 90089-03-02-06 & 53429.28 & $\phantom{0}$9808 \\
RXTE & 90089-03-02-09 & 53430.33 & $\phantom{0}$7536 \\
RXTE & 90089-03-02-10 & 53431.14 & $\phantom{0}$1376 & V \\
RXTE & 90089-03-02-12 & 53431.20 & $\phantom{0}$2896 & V \\
RXTE & 90089-03-02-14 & 53431.56 & $\phantom{0}$1248 & V \\
RXTE & 90089-03-02-11 & 53431.66 & $\phantom{0}$1360 & V \\
RXTE & 90089-03-02-15 & 53431.72 & $\phantom{00}$624 & V \\
RXTE & 90089-03-02-16 & 53432.12 & $\phantom{0}$1616 & V \\
RXTE & 90089-03-02-13 & 53432.18 & $\phantom{0}$1584 & V \\

  \multicolumn{5}{@{}l}{Outburst in 2007 December}\\
  RXTE & 93032-03-01-00 & 54426.00 & $\phantom{00}$560 \\
Swift & 00031030001 & 54427.69 & $\phantom{0}$2892 \\
RXTE & 93032-03-02-00 & 54427.81 & $\phantom{0}$2832 \\
RXTE & 93032-03-02-01 & 54429.84 & $\phantom{0}$2016 \\
RXTE & 93032-03-02-02 & 54431.61 & $\phantom{0}$2592 \\
RXTE & 93032-03-02-03 & 54433.05 & $\phantom{0}$1712 \\
RXTE & 93032-03-03-00 & 54434.24 & $\phantom{0}$1520 \\
Suzaku & 902003010 & 54434.48 & 34385 \\
RXTE & 93032-03-03-01 & 54435.50 & $\phantom{00}$863 \\
RXTE & 93032-03-03-02 & 54437.14 & $\phantom{00}$560 \\
RXTE & 93032-03-03-03 & 54438.17 & $\phantom{00}$768 \\
RXTE & 93032-03-03-04 & 54439.92 & $\phantom{0}$1760 \\
RXTE & 93032-03-04-01 & 54442.12 & $\phantom{00}$832 & VI \\
RXTE & 93032-03-04-02 & 54443.10 & $\phantom{00}$591 & VI \\
RXTE & 93032-03-04-00 & 54443.79 & $\phantom{00}$927 & VI \\
RXTE & 93032-03-04-04 & 54445.23 & $\phantom{0}$1200 & VII \\
RXTE & 93032-03-04-03 & 54446.33 & $\phantom{0}$1184 & VII \\

 \end{tabular}
\end{table}

\begin{table}
 \centering
 \addtocounter{table}{-1}
 \caption{continued}
 \label{tab:obsids2}
 \begin{tabular}{lcccc}
  Sat. & Observation & Starttime & Exp. & E \\ \hline\hline
  RXTE & 93032-03-04-06 & 54447.02 & $\phantom{0}$1360 & VII \\
RXTE & 93423-02-01-00 & 54449.18 & $\phantom{0}$1168 & VIII \\
RXTE & 93423-02-01-01 & 54451.98 & $\phantom{0}$1712 & VIII \\
RXTE & 93423-02-01-02 & 54454.87 & $\phantom{0}$1616 & VIII \\
RXTE & 93423-02-02-00 & 54457.16 & $\phantom{0}$1136 & VIII \\
\multicolumn{5}{@{}l}{Outburst in 2011 April}\\
RXTE & 96368-01-03-04 & 55647.35 & $\phantom{00}$655 & IX \\
RXTE & 96368-01-03-03 & 55648.39 & $\phantom{00}$671 & IX \\
RXTE & 96368-01-03-02 & 55649.62 & $\phantom{00}$688 \\
RXTE & 96368-01-03-01 & 55650.72 & $\phantom{00}$944 \\
RXTE & 96368-01-03-00 & 55651.97 & $\phantom{00}$768 \\
RXTE & 96368-01-02-06 & 55652.55 & $\phantom{0}$1104 \\
RXTE & 96368-01-02-05 & 55653.66 & $\phantom{00}$688 \\
RXTE & 96368-01-02-04 & 55654.50 & $\phantom{00}$719 \\
RXTE & 96368-01-02-03 & 55655.48 & $\phantom{00}$671 \\
RXTE & 96368-01-02-02 & 55656.59 & $\phantom{00}$704 \\
RXTE & 96368-01-02-01 & 55657.64 & $\phantom{00}$495 \\
RXTE & 96368-01-02-07 & 55658.23 & $\phantom{0}$1840 \\
RXTE & 96368-01-02-08 & 55658.37 & $\phantom{0}$2352 \\
RXTE & 96368-01-02-09 & 55658.43 & 13024 \\
RXTE & 96368-01-02-00 & 55658.72 & 15744 \\
RXTE & 96368-01-01-08 & 55659.15 & $\phantom{0}$1584 \\
RXTE & 96368-01-01-05 & 55659.55 & $\phantom{0}$6448 \\
RXTE & 96368-01-01-03 & 55659.67 & 16863 \\
RXTE & 96368-01-01-00 & 55660.32 & 14464 \\
RXTE & 96368-01-01-02 & 55660.65 & 17695 \\
RXTE & 96368-01-01-07 & 55661.37 & $\phantom{0}$2432 \\
RXTE & 96368-01-01-01 & 55661.43 & 14224 \\
RXTE & 96368-01-01-06 & 55661.96 & $\phantom{00}$655 \\

  \multicolumn{5}{@{}l}{Outburst in 2011 December}\\
  Swift & 00031030002 & 55915.83 & $\phantom{0}$1994 \\
Swift & 00031030003 & 55917.05 & $\phantom{0}$1801 \\
Swift & 00031030004 & 55918.78 & $\phantom{0}$1344 \\
Swift & 00031030005 & 55919.91 & $\phantom{0}$2032 \\

  \multicolumn{5}{@{}l}{After outburst in 2011 December }\\
  Swift & 00031030006 & 55932.08 & $\phantom{0}$1044 \\
Swift & 00031030007 & 55934.49 & $\phantom{0}$1196 \\
Swift & 00031030009 & 55938.77 & $\phantom{00}$969 \\
Swift & 00031030010 & 55940.42 & $\phantom{00}$986 \\
Swift & 00031030011 & 55955.80 & $\phantom{0}$1081 \\
Swift & 00031030012 & 55958.14 & $\phantom{0}$1039 \\
\multicolumn{5}{@{}l}{``Giant'' outburst in November 2012 }\\
Swift & 00031030013 & 56196.99 & $\phantom{0}$4533 \\
Swift & 00538290000 & 56244.91 & $\phantom{0}$3017 \\
Swift & 00031030014 & 56246.18 & $\phantom{0}$1982 \\
Swift & 00031030015 & 56248.25 & $\phantom{0}$1308 \\
Swift & 00031030016 & 56250.18 & $\phantom{0}$2021 \\
Suzaku & 907006010 & 56251.63 & $\phantom{0}$6521 \\
Swift & 00031030017 & 56252.59 & $\phantom{0}$2007 \\ \hline
 \end{tabular}
\end{table}

\begin{figure}
 \includegraphics[width=\columnwidth]{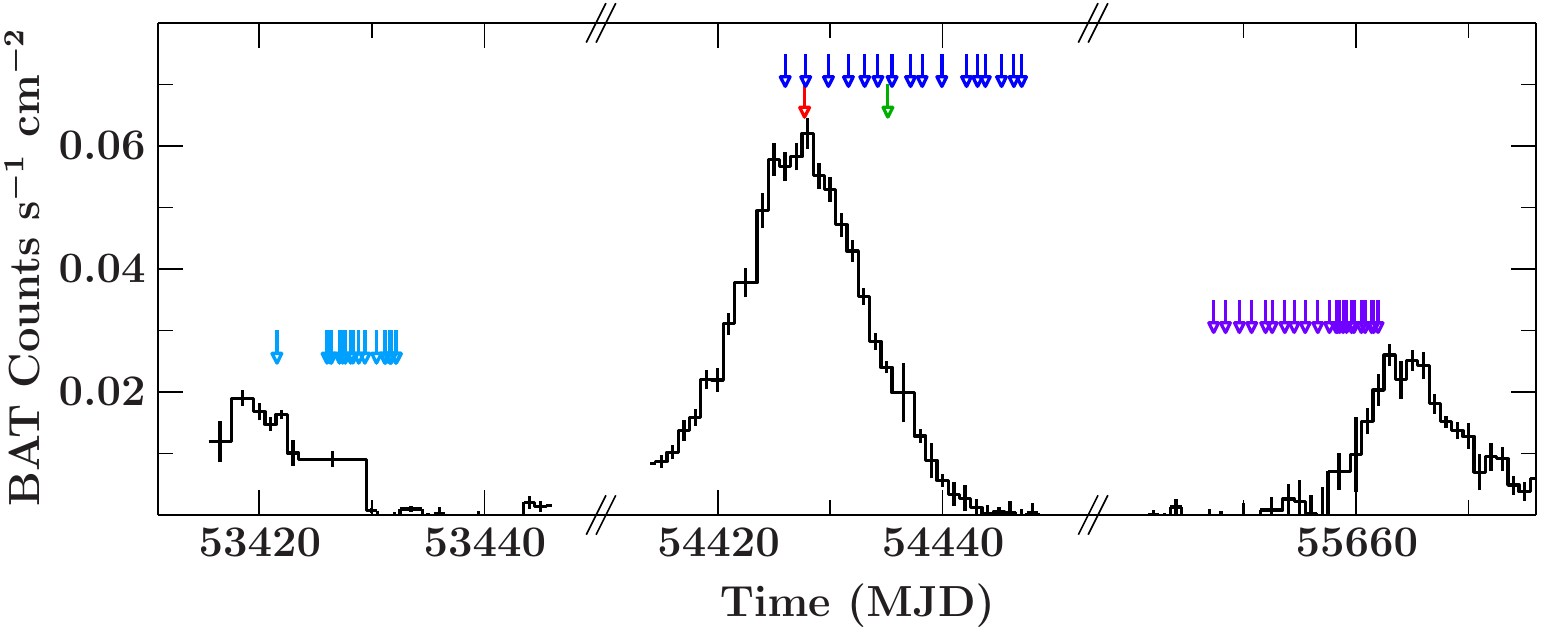}
 \caption{\Swift-BAT-light curve around the outbursts in 2005, 2007,
   and April 2011. The vertical arrows on top mark the start times of
   individual observations of \RXTE (light, dark blue, and purple),
   \Swift (red), and \Suzaku (green).}
 \label{fig:asm0507}

\vspace*{\baselineskip}

 \includegraphics[width=\columnwidth]{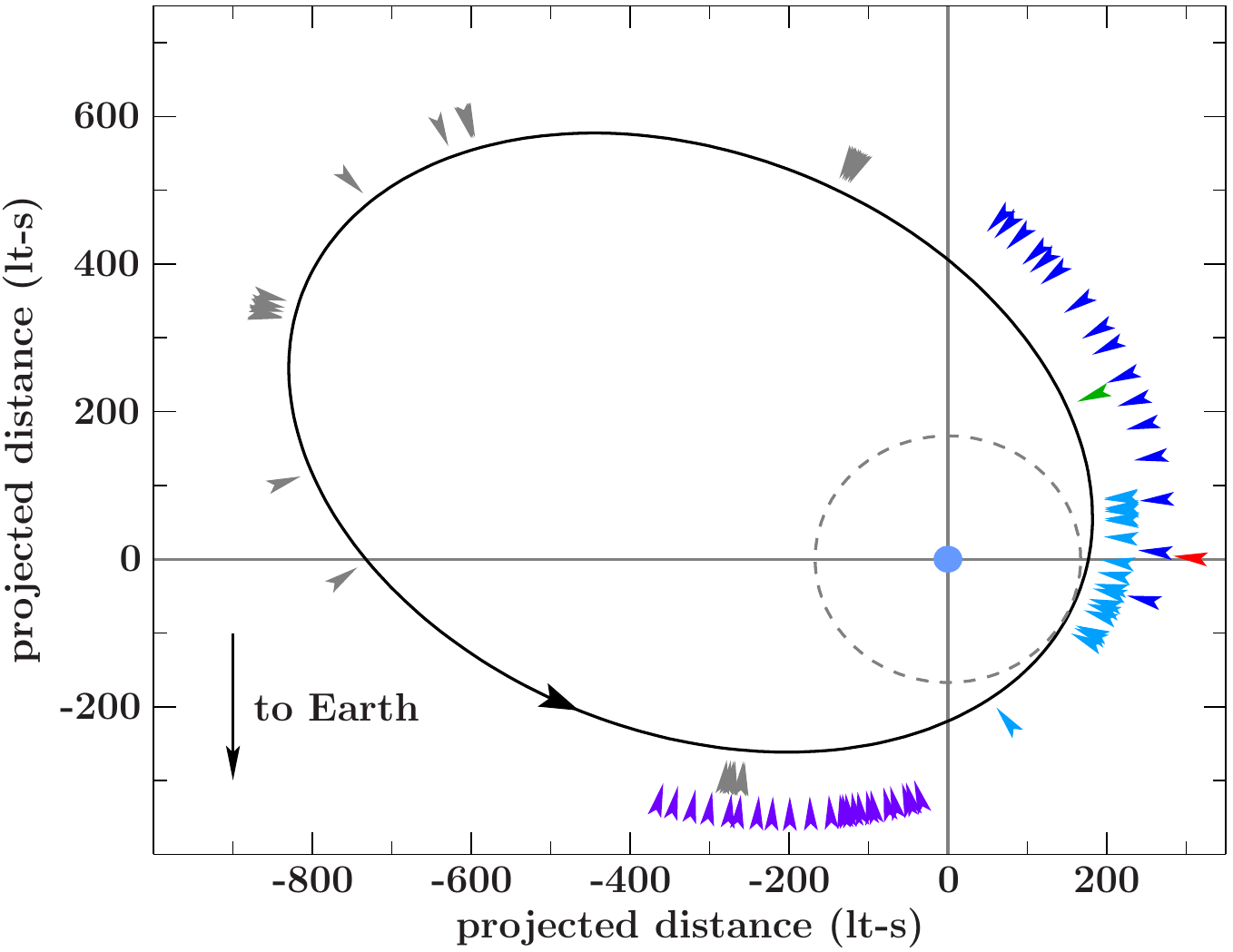}
 \caption{Projected orbital plane of the binary. The reference frame
   is centered on the optical companion, whose size is set to
   $7\,R_\odot$ \citep{coe2007a}. The dashed circle sketches a
   circumstellar disk of $72\,R_\odot$ as proposed by \citep{coe2007a}.
   Orbital phases of the observations from 2005, 2007, and 2011 April
   are marked by arrows. The colors are the same as in
   Fig.~\ref{fig:asm0507}. Grey arrows represent \RXTE observations
   done in quiescence (Obsids 20132-* and 20123-*). Orbital phases
   are calculated using the ephemeris given in Table~\ref{tab:orbit}.}
 \label{fig:orbitcover}
\end{figure}

Data from five outbursts of \gro are analyzed. The
earliest outburst studied here occurred in 2005 February and was well
covered by pointed \RXTE observations (Fig.~\ref{fig:asm0507}). In
2007 December, \gro underwent its strongest outburst during the
lifetime of the \RXTE-All Sky Monitor. The 2--10\,keV peak flux was
84(5)\,mCrab (Fig.~\ref{fig:asm0507}). Except for the rise, the whole
outburst was observed regularly by \RXTE. \Swift recorded the source
around maximum intensity and \Suzaku during its decay. During 2011
April we triggered \RXTE observations to cover the start of the third
outburst analyzed here. Later outbursts in 2011 December and 2012 were
covered by a few \Swift observations each. In addition, to measure the
diffuse emission from the region, data from an \RXTE observation
campaign in 1996/1997, performed while the source was off,
exists. Table~\ref{tab:obsids} lists a log of the observations.

The binary orbital phases of the 2005, 2007, and 2011 April
observations are shown in Fig.~\ref{fig:orbitcover}. Outbursts
occurred around the periastron of the orbit, where mass accretion
from the circumstellar disk is possible.

In the following the instruments used in the analysis of these
outbursts are described. All spectral and timing modeling was performed
with the \textit{Interactive Spectral Interpretation System}
\citep[ISIS,][version 1.6.2-7]{houck2000a}. Unless otherwise stated,
all error bars are given at the 90\% level for each parameter of
interest.

\subsection{RXTE} \label{sec:data:rxte}

The Proportional Counter Array \citep[PCA,][]{jahoda2006a} was one of
two pointed instruments onboard the \textsl{Rossi X-ray Timing
  Explorer} (\RXTE) and was sensitive for X-rays between 2 and 90\,keV.
The PCA consisted of five Proportional Counter Units (PCUs),
collimated to a field of view of $\sim$$1^\circ$. Unless stated
otherwise, only data from PCU2 were used, which was the best
calibrated PCU \citep{jahoda2006a}. Since the source can be weak away
from the peak of the outbursts, only data from the top layer of PCU2
were used. Data were reduced using \textsc{heasoft}~6.11 and using
standard data reduction pipelines \citep{wilms2006a}. Light curves for
the timing analysis were extracted using barycentered GoodXenon data
with 1\,s time resolution. For spectral analysis, 4.5\,keV--50\,keV
PCA data were used to avoid calibration issues around the Xenon L-edge.
Data from 10--20\,keV, 20--30\,keV,
30--40\,keV, and $>$40\,keV were rebinned by a factor of 2, 4, 6, and
8, respectively. In addition, systematic uncertainties of 0.5\% were
added to the spectra \citep{jahoda2006a}. To compensate for the few
percent uncertainty of the PCA's background model, the background was
scaled with a multiplicative factor.

Hard X-ray \RXTE data were obtained using the High Energy X-ray Timing
Experiment \citep[HEXTE,][]{rothschild1998a}. HEXTE consisted of two
clusters~A and B of four NaI/CsI scintillation detectors, alternating
between the source and background positions (``rocking''). In
2006~October, the rocking of cluster~A was turned off and the detector
was fixed in on-source position due to an anomaly in the rocking
mechanism. As no reliable background estimates are available,
cluster~A was not used for spectral analysis of the post 2006 data.
The rocking of cluster~B was switched off in 2010~April, thus no HEXTE
data was used for analysis during the 2011~April outburst.
20--100\,keV HEXTE data were used for spectral analysis and rebinned
by a factor of 2, 3, 4, and 10 in the energy intervals 20--30\,keV,
30--45\,keV, 45--60\,keV, and $>$60\,keV, respectively.

\begin{figure}
 \includegraphics[width=\columnwidth]{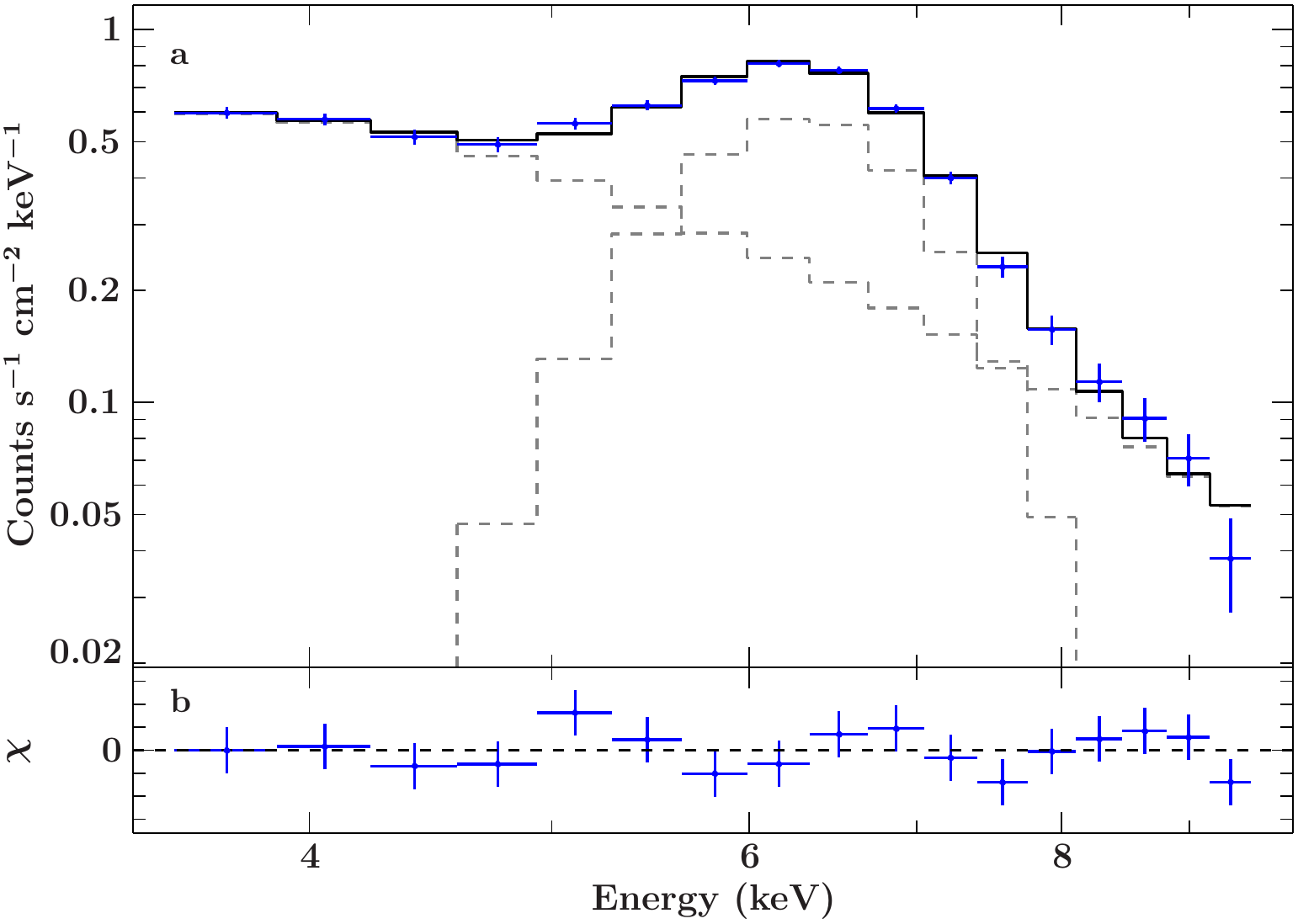}
 \caption{Galactic Ridge emission in the region around \gro as
   measured by \RXTE-PCA during a quiescent state of the source.
   \textbf{a)} The spectrum shows clear fluorescence emission of iron,
   which can be modelled by a broad Gaussian centered at 6.4\,keV, and
   a bremsstrahlung component. \textbf{b)} Residuals of the model.}
  \label{fig:galridgepca}
\end{figure}

The analysis of the \gro data is complicated due to the low Galactic
latitude of the source ($b=-1\fdg{}827$). At such low latitudes, the
cumulative effect of the Galactic ridge emission in the
${\sim}1^\circ$ radius field of view of the PCA is strong enough to
lead to visible features in the X-ray spectrum. When comparing
simultaneous PCA measurements with those from imaging instruments,
these features mainly show up as excess emission in the Fe K$\alpha$
band around 6.4\,keV and a slight difference in the continuum shape
\citep[e.g.,][]{mueller2012b}. The Galactic ridge emission is believed
to originate from many, unresolved X-ray binaries in the field of view
\citep{revnivtsev2009a} and can be empirically modelled by the sum of
a bremsstrahlung continuum and iron emission lines from neutral,
helium-like, and hydrogenic iron \citep{ebisawa2007a,yamauchi2008a}.
Due to the energy resolution of the PCA, the three lines merge into a
blended emission line.

The Galactic ridge emission shows a strong spatial variability.
Fortunately, \gro was observed by \RXTE in 1996 and 1997 during a
quiescent state (proposal IDs 20132 and 20123). The PCA light curves
from these observations do not show any evidence for pulsed emission
between 91 and 96\,s. Figure~\ref{fig:galridgepca} shows the 181\,ks
spectrum accumulated from these observations. In the spectral
analysis, this component was modeled as the sum of a bremsstrahlung
continuum and three Fe K lines and is only present in the PCA data
(the ridge emission is too soft to influence HEXTE).

\subsection{Swift} \label{sec:data:swift}

For the analysis of the \Swift observations data from its X-ray
Telescope's CCD detector, XRT, were used \citep{gehrels2004a}. The
1.5--9\,keV spectra were extracted in the ``Windowed Timing mode'', a
fast mode that prevents pileup for sources up to 600\,mCrab. Only the
observations after the 2011 December outburst and obs. 00031030013
before the giant outburst of 2012 October/November were extracted in
the ``Photon Counting mode''. Source spectra were obtained from a $1'$
radius circle centered on the source, or an annular circular region if
pileup was present. Background spectra were accumulated from two
circular regions off the source position with radii up to
$\sim$$1\farcm5$. Unless stated otherwise, spectral channels are added
up until a minimum signal-to-noise ratio of 20 is achieved. The light
curve was extracted with 1\,s time resolution.

\subsection{Suzaku} \label{sec:data:suzaku}

Similar to \Swift-XRT, the X-ray Imaging Spectrometer (XIS) mounted on
the \Suzaku satellite uses Wolter telescopes to focus the incident
X-rays \citep{koyama2007a}. After correcting the spacecraft attitude
\citep{nowak2011a} ``Normal Clock Mode'' 1--9\,keV data in 1/4 window
from front illuminated CCDs XIS0 and XIS3 and from the back
illuminated XIS1 were accumulated from annular circular extraction
regions with $1\farcm5$ radius. CCD areas with more than 4\% pileup
\citep{nowak2011a} were excluded. Depending on source brightness, the
circular exclusion region had a radius of typically $0\farcm5$. The
energy ranges 1.72\,keV--1.88\,keV and 2.19\,keV--2.37\,keV were
ignored due to calibration issues around the Si- and Au-edges
\citep{nowak2011a}. For the back illuminated XIS1, the energy range in
between the edges (1.88\,keV--2.19\,keV) was ignored as well due to
significant discrepancies with respect to the front illuminated XIS0
and XIS3. The spectral channels of XIS were added such that a combined
signal-to-noise ratio of 40 was achieved. For timing analysis,
XIS3-light curves with 2\,s resolution were used.

\Suzaku's collimated hard X-ray detector
\citep[HXD;][]{takahashi2007a} can be used to detect photons from
10\,keV up to 600\,keV. It consists of two layers of silicon PIN
diodes for energies below 50\,keV above phoswich counters sensitive
$>$57\,keV (GSO). Spectra from the energy ranges of 12--50\,keV (PIN)
and 57--100\,keV (GSO) were used. The PIN energy channels were binned to
achieve a signal-to-noise ratio of 20. For the GSO spectrum, a binning
factor of 2 was applied to channels between 60\,keV and 80\,keV and of
4 for higher energies.

Note that the pile up in the 2007 XIS data described above was not
taken into account in the previous analysis of these data by
\citet{naik2011a}.
If not excising the high count region of the core of the
point spread function, photons which arrive in the same or adjacent
pixels are misinterpreted by the detection chain as one photon of
larger energy. If the joint XIS/HXD spectra are then modeled, this
spectral distortion results in the apparent appearance of soft
spectral components claimed by \citet{naik2011a}. These components are
therefore not present in the real source spectrum.

\section{The Orbit of \gro} \label{sec:timing}

\begin{table}
 \centering
 \caption{Pulse periods and orbital parameters as determined by
   \cite{coe2007a} and in this paper. The parameters $e$,
   $a \sin{i}$, and $\omega$ remain unchanged in the analysis
   presented here. Uncertainties of the new values are at the 90\%
   confidence level ($\chi^2_\mathrm{red}=1.04$ for 1024 dof).}
 \label{tab:orbit}
 \renewcommand{\arraystretch}{1.3}
 \begin{tabular}{ll}
   \hline
   \multicolumn{2}{l}{\citet{coe2007a}} \\
   \hline
   Eccentricity               & $e=0.68(2)$ \\
   Projected semi major axis  & $a \sin{i}=530(60)$\,lt-s \\
   Longitude of periastron    & $\omega= -26(8)^\circ$ \\ 
   Orbital period             & $P_\mathrm{orb}=247.8(4)$\,d \\
   Time of periastron passage & $\tau= \mathrm{MJD}\,49189.8(5)$ \\
   \hline
   \multicolumn{2}{l}{This paper} \\ 
   \hline
   Orbital period                & $P_\mathrm{orb}=249.48^{+0.04}_{-0.04}$\,d \\
   Time of periastron passage    & $\tau=\mathrm{MJD}\,54424.71^{+0.20}_{-0.16}$ \\
   Spin period during 2005       & $P_\mathrm{2005}=93.67928^{+0.00010}_{-0.00009}$\,s \\
   Spin period during 2007       & $P_\mathrm{2007}=93.71336^{+0.00017}_{-0.00022}$\,s \\
   Spin-up before MJD\,54434.4819 & $\dot{P}_\mathrm{2007}=-0.61^{+0.24}_{-0.22}\times 10^{-9}\,\mathrm{s}\,\mathrm{s}^{-1}$ \\
                                  & $\ddot{P}_\mathrm{2007}=3.38^{+0.07}_{-0.16}\times 10^{-14}\,\mathrm{s}\,\mathrm{s}^{-2}$ \\ \hline

 \end{tabular}
\end{table}

\subsection{Orbit Determination Using Pulse Arrival Time
  Measurements} \label{sec:timing:arrmethod} 

Between the 2007 outburst and the reported periastron passage of
the original orbit determination by \citet{coe2007a} using
\CGRO-BATSE, \gro orbited its companion 21 times. The accumulated
uncertainty of the ephemeris of \citet{coe2007a} during this time
is about 8.5\,d (0.03\,$P$), or almost as long as a
typical outburst of the source. Not correcting for this uncertainty
in the analysis of the pulsar's spin could introduce false
spin-up/spin-down phases.
 
In order to derive the orbit of \gro the method of pulse arrival
timing was used \citep[see also][and references
therein]{deeter1981a,boynton1986a}. Assuming a neutron star with a
slowly varying pulse period, the arrival time $t(n)$ of the $n$th
pulse is given using a Taylor expansion of the pulse ephemeris up to
the third order,
\begin{multline} \label{eq:pulstime}
  t(n) =  t_0 + P_0 n + \frac{1}{2!} P_0 \dot{P} n^2 + \frac{1}{3!} \left(P_0^2 \ddot{P} + P_0 \dot{P}^2\right) n^3 \\
      + \frac{a_x \sin{i}}{c} F(e,\omega,\tau,\theta)
\end{multline}
where $P_0$ is the pulse period in the rest frame of the neutron star
at the reference time $t_0$, $a_x \sin{i}$ is the projected semimajor
axis of the neutron star's orbit and $\dot{P}$ and $\ddot{P}$ are the
changes in the spin period. The function $F$ describes the time delay
due to an orbit of eccentricity $e$, longitude of periastron $\omega$,
time of periastron passage $\tau$, and mean anomaly $\theta$. The
latter is connected to the orbital period $P_\mathrm{orb}$ via $\theta = 2
\pi (t - \tau)/P_\mathrm{orb}$ \citep[see also][]{kelley1980a,nagase1989a}.
Note that for reasons of computational speed and numerical accuracy,
$t(n)$ should be evaluated using a \citet{horner1819a} scheme.

To determine the pulse arrival times, a template pulse profile
determined by folding part of an observation with a locally determined
pulse period \citep{leahy1983a} was correlated with the observed light
curves. Note that an individual profile was created for each instrument
used, so that changes in the profile shape caused by time evolution or
by different energy ranges and responses of the detectors were taken
into account. 

Having found the arrival times, the difference between measured and
predicted arrival time according to Eq.~\eqref{eq:pulstime} was
minimized by varying the orbital parameters as well as $P_0$,
$\dot{P}$, and $\ddot{P}$ using a standard $\chi^2$ minimization
algorithm. Since the template pulse profile depends on the orbit
correction and pulse ephemeris, i.e., on the fit parameters, the whole
process was iterated until convergence was reached.

\begin{figure}
 \includegraphics[width=\columnwidth]{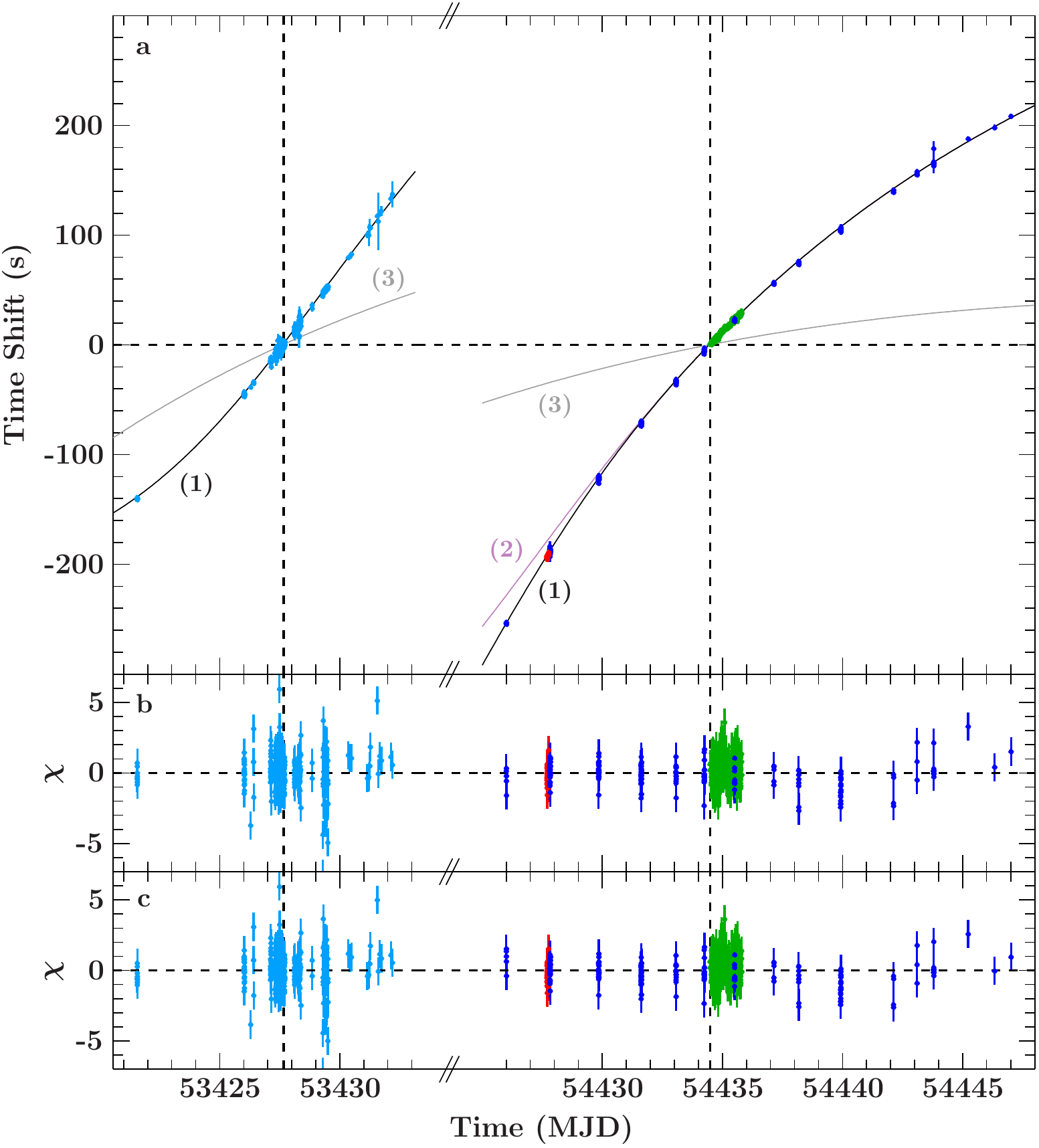}
 \caption{\textbf{a)} Time shifts relative to a constant
   pulse period for the 2005 and 2007 outbursts from \RXTE- (light
   blue: 2005, dark blue: 2007), \Swift- (red) and \Suzaku-data
   (green). The arbitrary reference time for the pulse ephemeris is
   indicated by the vertical dashed lines. The gray curve (3) shows
   the effect of the orbital correction using the orbit of
   \citet{coe2007a} on the arrival times. The best-fit pulse ephemeris
   based on the revised orbital parameters of Table~\ref{tab:orbit}
   are shown in black with (1) and in purple without a spin-up (2).
   \textbf{b)} Residuals of the model without taking the times of
   maximum source flux during outbursts into account. \textbf{c)}
   Residuals of the best-fit model (1).}
 \label{fig:arrtimes}
\end{figure}

\subsection{The Orbit of \gro} \label{sec:timing:analysis}

\begin{figure}
 \includegraphics[width=\columnwidth]{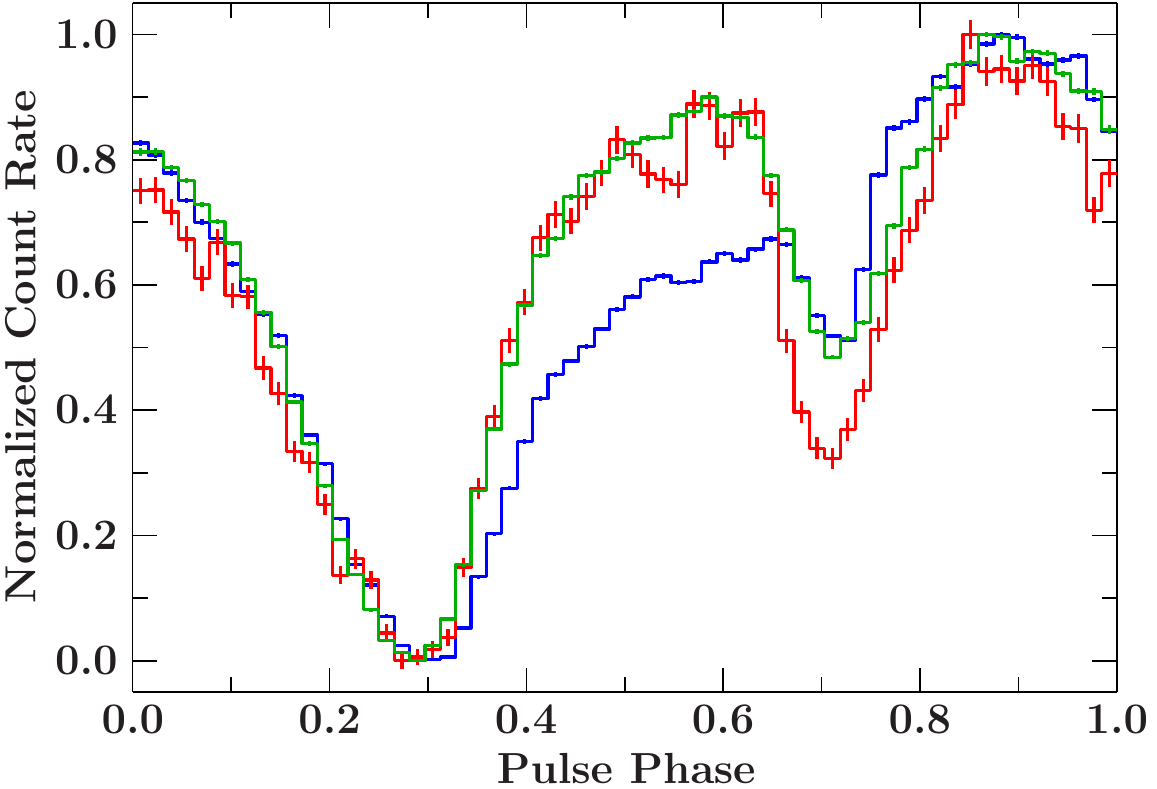}
 \caption{The \RXTE-PCA (blue), \Swift-XRT (red), and \Suzaku-XIS3 pulse
   profiles (green). The profiles are folded on the first
   time bin of the extracted XIS3 lightcurve (MJD~54434.4819). Due
   to a comparable sensitive energy range the profiles as detected by
   \Swift-XRT (0.2--10\,keV) and \Suzaku-XIS3 (0.2--12\,keV) are similar in
   shape, while the secondary peak is much more prominent at higher energies
   as seen in \RXTE-PCA (2--60\,keV).
   In order to show the good agreement between the \Swift and \Suzaku data,
   profiles are shown with 64 bins instead of the 32 bins used for the 
   arrival time analysis.}
 \label{fig:profiles}
\end{figure}

To determine the orbit of \gro all data available from the 2005 and
2007 outbursts were used. Template pulse profiles have 32\,phase bins.
During the 2005 outburst a period of $P_\mathrm{obs}=93.698(7)$\,s was
found, while for the 2007 one the period is
$P_\mathrm{obs}=93.7369(12)$\,s. The reference time needed to
determine the time shifts was set to the first time bin of the
\Suzaku-light curve (MJD\,54434.4819) for data taken in 2007, and to
the first time bin of \RXTE-observation 90089-03-02-00 (MJD
53427.6609) for the 2005 outburst. See Fig.~\ref{fig:profiles} for example 
pulse profiles.

Figure~\ref{fig:arrtimes} shows the difference between the measured
pulse arrival times and a pulse ephemeris which assumes constant pulse
periods with values given above. Calculating the arrival time using
the ephemeris of \citet[][see Table~\ref{tab:orbit}]{coe2007a} did not
result in a good description of the data (Fig.~\ref{fig:arrtimes},
gray lines). In principle, the remaining difference could be explained
by a very large spin-up and spin-down. If the difference is explained
in this way, the resulting orbital parameters are, however,
inconsistent with the phasing of the outbursts seen with all sky
monitors (see Sect.~\ref{sec:timing:asmbat} below). The only orbital
parameters that can result in the difference between the gray line in
Fig.~\ref{fig:arrtimes} and the data are the epoch of periastron
passage, $\tau$, and the orbital period, $P_\mathrm{orb}$. A change in
eccentricity, $e$, $\omega$, or $a\sin i$ shifts arrival times in a
way that would move the lines in Fig.~\ref{fig:arrtimes} up and down
or stretch the whole figure. Neither change results in a good
description of the data. Holding all parameters except $P_0$, $\tau$,
and $P_\mathrm{orb}$ fixed, and performing a $\chi^2$ minimization
then results in a good description of all arrival times
(Fig.~\ref{fig:arrtimes}, black and purple curves).

Close inspection of the best fit revealed, however, a deviation
between the best fit and the data from the early measurements during
the 2007 outburst, which were taken during the maximum of the
outburst. This discrepancy at high luminosities during the 2007
outburst can be explained by assuming that during this phase the
pulsar underwent a spin up, i.e., $\dot{P}\ne
0\,\mathrm{s}\,\mathrm{s}^{-1}$ and $\ddot{P}\ne
0\,\mathrm{s}\,\mathrm{s}^{-2}$, between MJD\,54426 and MJD\,54434.
This approach is not unreasonable since high luminosity phases are
coincident with a high angular momentum transfer onto the neutron
star, which has been detected in other transient X-ray binaries as
well, e.g., \object{A0535+26} \citep{camero-arranz2012a}.

Using the resulting orbital parameters, a first inspection of the
orbital phases where the outbursts of \gro are detected in \RXTE-ASM
and \Swift-BAT showed that the times of maximum source flux are
consistent with a single orbital phase (see Sect.~\ref{sec:timing:asmbat}). 
To enhance the orbital
parameter precision further, especially the orbital period,
the times of maximum source flux were fitted
simultaneously with the arrival times data. The residuals are very
similar to those found previously (compare residual panels b and c
of Fig.~\ref{fig:arrtimes}).

The best fit parameters of the final orbital solution and pulse period
ephemeris are presented in Table~\ref{tab:orbit}. Note that the
simultaneous fit of the 2005 and 2007 arrival times data as well as the
outburst times in ASM and BAT allow to break the correlation between
the orbital period and the time of periastron passage which is
generally found when studying pulse arrival times from one outburst only.

\subsection{ASM- and BAT-analysis} \label{sec:timing:asmbat}

The orbital period of 249.48(4)\,d found by the combined pulse arrival
times and outburst times analysis in the previous section differs by
more than $4\sigma$ from the value found by
\citet[][$P_\mathrm{orb}=247.8(4)$\,d]{coe2007a} and by slightly more
than $1\sigma$ from the orbital period found by
\citet[][$P_\mathrm{orb}=248.9(5)$\,d]{levine2006a} from the \RXTE-ASM
light curves. Since outbursts occur during each periastron passage, as
a consistency check $P_\mathrm{orb}$ can also be determined by
measuring the time of peak flux from the available \RXTE-ASM and
\Swift-BAT data only. The average distance between outburst peaks is
249.7(4)\,d, which is in agreement with the results of the arrival
times analysis of Sect.~\ref{sec:timing:analysis} and
\citet{levine2006a}.

Having confirmed the updated orbital period, it is possible to compare
the time of maximum source flux with the periastron passages predicted
by the revised orbital solution. With the exception of the bright 2007
outburst (see below), the results displayed in
Fig.~\ref{fig:outburst_phase} show that outburst maxima are consistent
with a mean orbital phase of $\phi_\mathrm{orb} = -0.0323(17)$ where
the uncertainty includes the uncertainties of the orbital parameters.
Thus, \gro reaches maximum luminosity during outburst very close to,
but significantly before, periastron.

Following the improved orbital parameters and the detected phase shift
between the peak flux and the periastron passage, the date of the
highest flux of \gro during an outburst can be predicted:
\begin{equation}
 T_\mathrm{max} = \mathrm{MJD}\,54416.65 + n \times 249.48
\end{equation}
where $n$ is the number of orbits since the outburst in 2007. The
uncertainty of $T_\mathrm{max}$ is about 3\,d. It is mainly due to the
scattering of the outburst times (Fig.~\ref{fig:outburst_phase}). A
successful prediction of the 2011 April outburst resulted in the \RXTE
observations during the rise of this outburst, which are analysed in
this work.

The ASM analysis shows that the 2007 outburst of \gro, the brightest
outburst seen by \RXTE-ASM, was delayed by $\sim11$\,days compared to
the standard ephemeris. The time of maximum luminosity from this
outburst was therefore excluded from the analysis above. However,
as the typical duration of one outburst of \gro is $\sim$14\,days
($0.056\,P_\mathrm{orb}$), even this outburst is clearly connected
with a periastron passage of the neutron star.

\begin{figure}
 \includegraphics[width=\columnwidth]{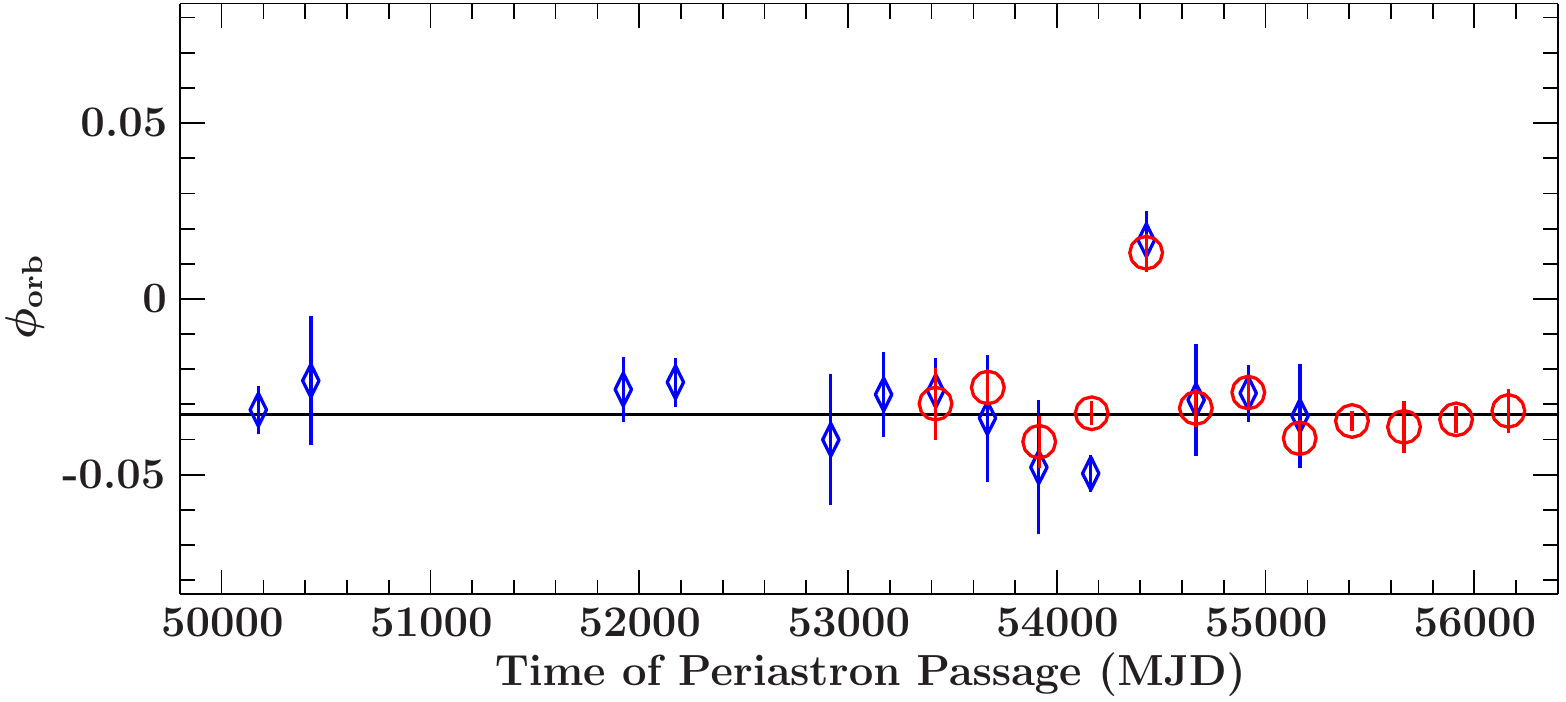}
 \caption{Orbital phases of each detected outburst of \mbox{\gro} in
   ASM (blue) and BAT (red), determined from the orbital parameters
   found by the arrival times analysis (see text). The solid line
   shows the best-fit outburst phase of the ASM- and  BAT-data. The
   outlier is the 2007 December outburst.}
 \label{fig:outburst_phase}
\end{figure}

\section{Spectral Modelling} \label{sec:spectral}

\subsection{Continuum of \gro} \label{sec:spectral:cont}

The continuum emission of accreting neutron stars is produced in the
accretion columns over the magnetic poles of the compact object
\citep[][and references therein]{becker2007a}. Here, the kinetic
energy of the infalling material is thermalized and a hot spot forms
on the surface that is visible in the X-ray band as a black body.
A fraction of that radiation is then Compton upscattered in the
accretion column above the hot spot, forming a hard power law spectrum
with an exponential cutoff at around 50--150\,keV.

Although theoretical models now start to emerge which allow to model
the continuum emission directly \citep{becker2007a,ferrigno2009a},
these models are not yet self-consistent enough to describe the
spectra of all accreting neutron stars. For this reason, empirical
models are typically used (see \citealt{mueller2012b} and
\citealt{decesar2013a} for recent
summaries).

\begin{figure}
 \includegraphics[width=\columnwidth]{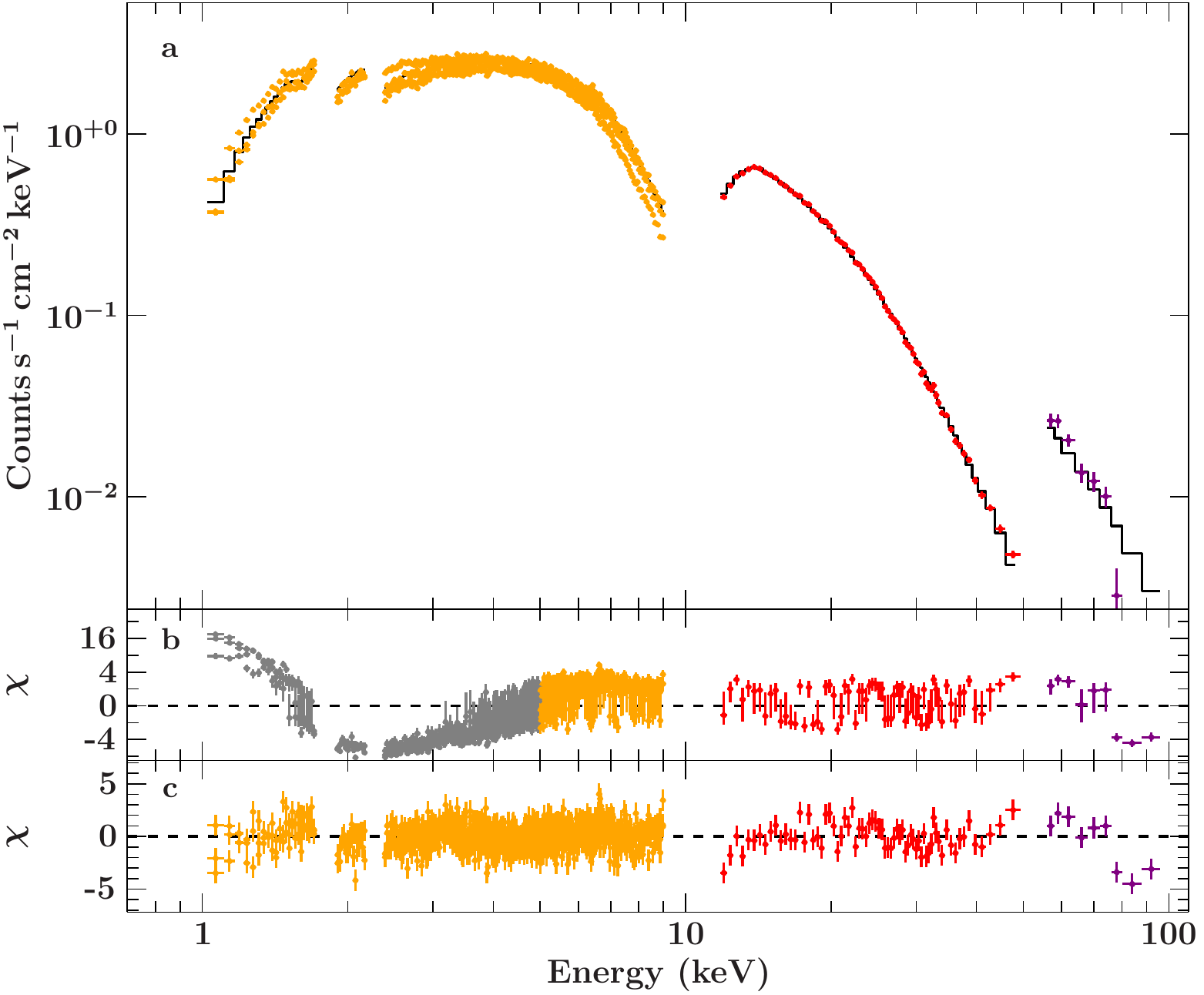}
 \caption{\textbf{a)} The 2007 \Suzaku-spectrum of \gro (XISs: orange,
   PIN: red, GSO: purple). Spectral channels are rebinned for display
   purposes. \textbf{b)} A pure \texttt{CUTOFFPL}-model fitted to data above
   5\,keV, to compare with \INTEGRAL-data \citep{coe2007a}, results in
   a poor description below that energy (gray residuals). \textbf{c)}
   Adding a black body and a narrow iron line, and taking absorption
   in the  interstellar medium into account, improves the fit to
   $\chi^2_\mathrm{red} \sim 1.12$ (1845\,dof).}
  \label{fig:specmodels}
\end{figure}

Due to the mentioned exponential cutoff of the spectrum, data from
instruments such as the \CGRO-OSSE and -BATSE mainly show the
exponential roll over above 20\,keV. As a result, the combined
\CGRO-OSSE and -BATSE spectrum (20--150\,keV) during the discovery
outburst of \gro in 1993 was initially described by a bremsstrahlung
continuum \citep{shrader1999a}. As discussed by \citet{shrader1999a}
for the joint 1993 \CGRO/\textsl{ASCA}-GIS (0.7--10\,keV) spectrum and
also by \citet{coe2007a} for \INTEGRAL data from the 2004 June
outburst, the soft X-ray spectrum cannot be described by a
bremsstrahlung continuum. Instead, a power law spectrum with an
exponential cutoff of the form
\begin{equation}
  \texttt{CUTOFFPL}(E) \propto E^{-\Gamma} e^{-E/E_\mathrm{fold}}
\end{equation}
was found to be a good description of the spectrum. Here $\Gamma$ is
the photon index and $E_\mathrm{fold}$ the folding energy.

\begin{figure}
 \includegraphics[width=\columnwidth]{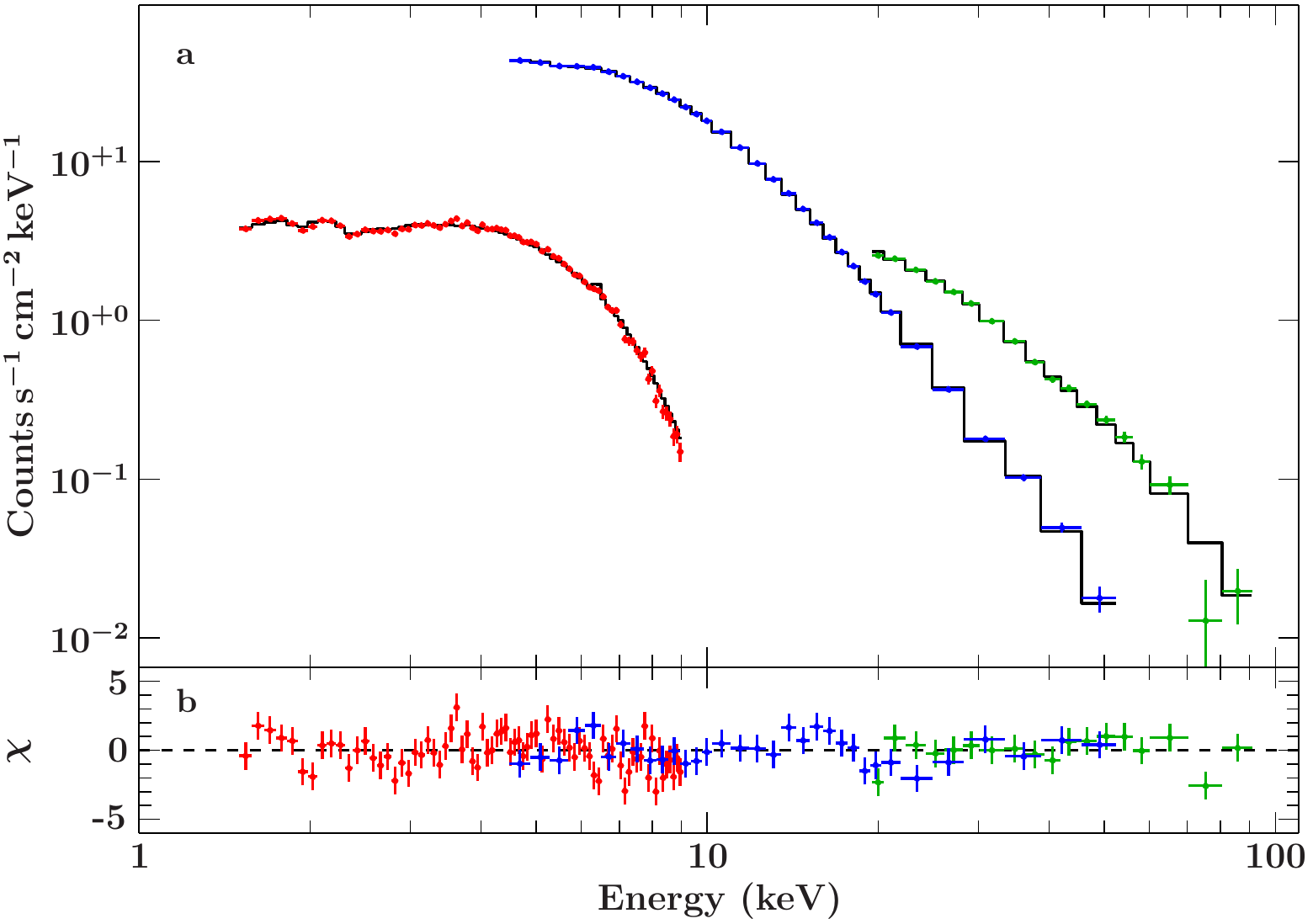}
 \caption{\textbf{a)} Quasi-simultaneous 2007 \Swift and \RXTE spectra
   of \gro (\Swift-XRT: red, \RXTE-PCA: blue, \RXTE-HEXTE: green).
   Spectral channels are grouped for display purposes. \textbf{b)}
   Residuals of a fit to the model given by Eq.~\ref{eq:model}.}
  \label{fig:combinedfits}
\end{figure}

Applying this model to the 2007 \Suzaku data gives a good description
of the spectrum above $\sim$5\,keV (Fig.~\ref{fig:specmodels}a). At
softer energies deviations are present. Note that these differences
are not the same as those caused by pile up discussed in
Sect.~\ref{sec:data:suzaku}. The broad band deviations are well
described by a black body spectrum modified by interstellar
absorption. A narrow iron K$\alpha$ fluorescence line at 6.4\,keV is
needed to achieve an acceptable fit with a reduced
$\chi^2_\mathrm{red}=1.14$
for 1896 degrees of freedom (dof; see Fig.~\ref{fig:specmodels}). 
The iron K$\beta$ fluorescence line at 7.056\,keV was included
with an equivalent width fixed to 13\% of the width of the K$\alpha$ line.
This additional K$\beta$ line is included in all spectral fits in this
paper. The $\chi^2$ can be further decreased to
$\chi^2_\mathrm{red}=1.11$ (1888 dof) by ignoring the GSO spectrum. The
reason are residuals above
75\,keV, which might be caused by the putative cyclotron line at
88\,keV \citep{shrader1999a}. The signal quality does not, however,
allow to constrain the cyclotron line parameters properly. See
Sect.~\ref{sec:spectral:crsf} for further discussion of that feature.
Adding the simultaneous \RXTE spectrum (93032-03-03-01) to
the \Suzaku data, and taking the Galactic ridge emission into account,
does not appreciably change the best fit parameters.
Table~\ref{tab:combinedparams}\footnotemark[2] lists the final best fit
parameters.
Applying the same model to the quasi-simultaneous \Swift-XRT,
\RXTE-PCA and \RXTE-HEXTE data (\RXTE obsid 93032-03-02-00) from the
same 2007 outburst also gives a good description of the spectrum (see
Fig.~\ref{fig:combinedfits} and Table~\ref{tab:combinedparams}\footnote{
  The parameters of the GRE are
  determined by a combined fit of all data from all outbursts and
  therefore listed in Table~\ref{tab:alldatafitpars} (see
  Sect.~\ref{sec:spectral:evolution}).}).

\begin{table}
   \caption{Best-fit continuum parameters of \gro determined from the
     simultaneous 2007 \Suzaku and \RXTE data
     ($\chi^2_\mathrm{red} = 1.15$, 1935\,dof) and the quasi simultaneous
     2007 \Swift-XRT and \RXTE data ($\chi^2_\mathrm{red} = 1.15$,
     160\,dof). The fit takes Galactic ridge emission applied to
     \RXTE-PCA into account. The given uncertainties are at
     the 90\%-confidence limit. See Table~\ref{tab:alldatafitpars} for
     the parameters of the Galactic ridge emission
     model.}\label{tab:combinedparams} 
\renewcommand{\arraystretch}{1.3}

 \begin{tabular}{llp{10mm}p{10mm}l}
   component & & \Suzaku \RXTE & \Swift \RXTE & unit \\
   \hline\hline
   \texttt{TBnew} & $N_\text{H}$ & $1.523^{+0.029}_{-0.029}$ & $1.56^{+0.17}_{-0.17}$ & $10^{22}\,\text{cm}^{-2}$ \\
\texttt{CUTOFFPL}\tablefootmark{a} & $\Gamma$ & $0.522^{+0.024}_{-0.024}$ & $0.57^{+0.07}_{-0.07}$ &  \\
 & $E_\text{fold}$ & $15.6^{+0.5}_{-0.4}$ & $16.1^{+0.8}_{-0.8}$ & keV \\
 & $F_\text{PL}$ & $1.982^{+0.029}_{-0.029}$ & $3.97^{+0.08}_{-0.08}$ & $10^{-9}$\,erg\,s$^{-1}$\,cm$^{-2}$ \\
\texttt{BBODY}\tablefootmark{a} & $kT$ & $1.854^{+0.025}_{-0.025}$ & $1.86^{+0.06}_{-0.05}$ & keV \\
 & $F_\text{BB}$ & $0.501^{+0.016}_{-0.016}$ & $0.77^{+0.08}_{-0.08}$ & $10^{-9}$\,erg\,s$^{-1}$\,cm$^{-2}$ \\
iron line      & $E$ & \multicolumn{2}{c}{6.4 (fix)} & keV \\
& $\sigma$ & \multicolumn{2}{c}{$10^{-6}$ (fix)} & keV \\
 & $W$ & $23.5^{+2.5}_{-2.5}$ & $41^{+10}_{-10}$ & eV \\
constants\tablefootmark{b} & $c_\text{HEXTE}$ & $0.84^{+0.04}_{-0.04}$ & $0.853^{+0.019}_{-0.019}$ &  \\
 & $c_\text{XRT}$ & - & $0.809^{+0.010}_{-0.010}$ &  \\
 & $c_\text{XIS0}$ & $0.804^{+0.007}_{-0.006}$ & - &  \\
 & $c_\text{XIS1}$ & $0.845^{+0.007}_{-0.007}$ & - &  \\
 & $c_\text{XIS3}$ & $0.801^{+0.006}_{-0.006}$ & - &  \\
 & $c_\text{PIN}$ & $0.929^{+0.012}_{-0.010}$ & - &  \\
 & $c_\text{GSO}$ & $1.15^{+0.10}_{-0.10}$ & - &  \\
 & $b_\text{PCA}$ & $0.93^{+0.05}_{-0.05}$ & $0.96^{+0.04}_{-0.04}$ &  \\

   \hline
 \end{tabular}
 \tablefoot{
 \tablefoottext{a}{$F_\mathrm{PL}$ and $F_\mathrm{BB}$ are unabsorbed
 fluxes, $F_\mathrm{PL}$ is the power law flux in 15--50\,keV,
 $F_\mathrm{BB}$ is the bolometric black body flux.}
 \tablefoottext{b}{Detector flux calibration constants, $c$, are given
   relative to the \RXTE-PCA; the PCA background is multiplied by
   $b_\mathrm{PCA}$.}
}
\end{table}

In summary, the broad band spectrum of \gro can be described by a
model of the form
\begin{align} \label{eq:model}
 \nonumber F_\mathrm{ph, model}(E)   &= \texttt{TBnew} \times (\texttt{CUTOFFPL}+\texttt{BBODY} +
 \mathrm{Fe}_\mathrm{6.4\,keV} + \\ & \mathrm{Fe}_\mathrm{6.67\,keV}) + [\mathrm{GRE}] \\
\intertext{where $\mathrm{Fe}_\mathrm{6.4\,keV}$ is a Gaussian
  emission line, \texttt{BBODY} describes a black body, and where
  the  Galactic ridge emission is}
 \mathrm{GRE} &= \texttt{TBnew} \times (\texttt{BREMSS} +
 \mathrm{Fe}_\mathrm{blend})
 \label{eq:modelgre}
\end{align}
where \texttt{TBnew} is a revised version of the absorption model of
\citet[][see also \citealt{hanke:09a}]{wilms2000a}, using the
abundances of \citet{wilms2000a}, and where \texttt{BREMSS} is a
bremsstrahlung spectrum. The best-fit hydrogen column densities,
$N_\mathrm{H}$, are consistent between both fits. They are also in agreement
with the foreground absorption found in 21\,cm surveys
($N_\mathrm{H}=1.35^{+0.21}_{-0.09} \times 10^{22}\,\mathrm{cm}^{-2}$,
\citealt{kalberla:05a}, and $N_\mathrm{H}=1.51^{+0.38}_{-0.02} \times
10^{22}\,\mathrm{cm}^{-2}$, \citealt{dickey:90a}), as well as with the
hydrogen column density of $1.49\times10^{22}\,\mathrm{cm}^{-2}$
obtained from the interstellar reddening of
$E^\mathrm{is}(B-V)=1.79(5)$\,mag towards \gro \citep[][converted to
$N_\mathrm{H}$ as outlined by \citealt{nowak2012a}]{riquelme2012a}.

\subsection{Combined parameter evolution} \label{sec:spectral:evolution}

A closer inspection of the best fits from both multi-satellite
campaigns in Table~\ref{tab:combinedparams} shows that most of the
continuum parameters appear to be constant within their uncertainties.
The only significant parameter change between both models is a change
in the power law flux by a factor of $\sim$2 due to the change of flux
of the source over the outburst.

Fitting all \RXTE observations of the 2007 outburst with the basic
model of Eq.~\eqref{eq:model} gives acceptable $\chi^2$-values for all
observations. With the exception of the fluxes of the two continuum
components, other spectral parameters appear to be constant within
their error bars, although some scatter is still visible. Due to the
lower sensitivity of \RXTE compared to \Suzaku, however, this scatter
could be purely statistical in nature.

\begin{figure}
 \centering
 \includegraphics[width=.8\columnwidth]{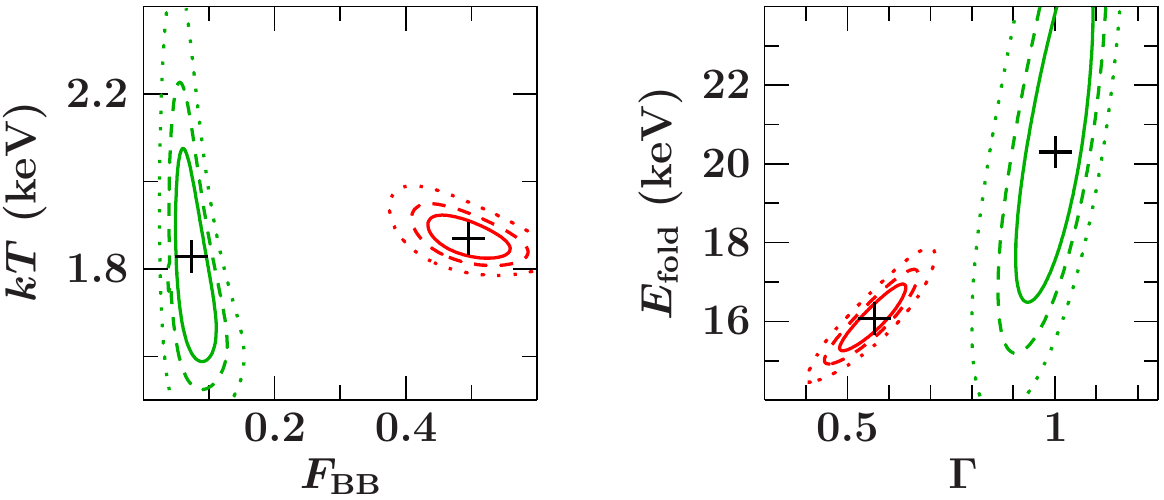}
  \caption{Contour maps between several continuum parameters during
    the 2007 outburst. Each map is calculated from RXTE data near
    maximum luminosity (red; obs. 93032-03-02-00) and at the end of
    the outburst (green; obs. 93032-03-03-04). The solid line
    represents the $1\sigma$ contour, the dashed line $2\sigma$ and
    the dotted line $3\sigma$. The black body flux, $F_\mathrm{BB}$, is
    given in units of
    $10^{-9}\,\mathrm{erg}\,\mathrm{s}^{-1}\,\mathrm{cm}^{-2}$.}  
  \label{fig:confmaps}
\end{figure}

To illustrate the range of parameter
variability, Fig.~\ref{fig:confmaps} shows contour maps between
several continuum parameters calculated from \RXTE data during maximum
luminosity in 2007 (93032-03-02-00) and at the end of the 2007 outburst
(93032-03-03-04).
The contour map on the left demonstrates that the black body flux
$F_\mathrm{BB}$ changes significantly over the outburst, while its
temperature $kT_\mathrm{BB}$ can be described by the same values.
Keeping this value fixed during both observations at a consistent
value improves the uncertainty on the remaining parameters
significantly. In particular, a contour map of $\Gamma$ and
$E_\mathrm{fold}$ reveals a strong correlation. Similar to the
black body temperature and the break energy, the folding energy
seems to be consistent with a constant value during the outburst,
while the photon index changes.

Inspection of the confidence maps reveals that the only spectral
parameters for which significant changes are seen in individual
spectral fits appear to be the power law flux, $F_\mathrm{PL}$, the
photon index $\Gamma$, and the black body flux $F_\mathrm{BB}$. To
constrain these parameters well and to reveal the evolution of the
remaining parameters, all datasets available during the 2007 outburst
(see Table~\ref{tab:obsids}) including the simultaneous \Swift and
\Suzaku are fit simultaneously with the spectral continuum of
Eq.~\eqref{eq:model} (and the Galactic ridge). In these fits the
continuum parameters $N_\mathrm{H}$,
$E_\mathrm{fold}$, and $kT$, and the flux calibration constant
$C_\mathrm{HEXTE}$ are not allowed to vary between the different
observations. The free parameters of the simultaneous fit are the
fluxes of the spectral components, $F_\mathrm{PL}$ and
$F_\mathrm{BB}$, the photon index, $\Gamma$, the iron line equivalent
width $W$, and the PCA background scaling factor $b_\mathrm{PCA}$.
Since the initial fit showed that the equivalent width of the iron line
$W$ does not change significantly with source flux, in a final fit this
parameter is also assumed to be the same for all observations.

The resulting model describes all 31 spectra of the 2007 outburst
with an $\chi_\mathrm{red}^2=1.12$ for 2623 dof. Adding all \RXTE-data
of the 2005 and 2011 April outbursts to
check for possible hysteresis effects and refitting shows that the
simplified model still gives a good description of all data if the
equivalent width of the iron line is allowed to change between
different outbursts (but remaining constant during each outburst).
With $N_\mathrm{H}=1.6\times10^{22}\,\mathrm{cm}^{-2}$, the soft X-ray
absorption is too low to be constrained by PCA-data only. For this
reason, changes of $N_\mathrm{H}$ between the outbursts cannot be
detected in the PCA data and the column density is held fixed for all
outbursts. The best-fit $\chi^2$ for this fit of 68 spectra is
$\chi_\text{red}^2=1.10$ for 3639 dof. Finally, to constrain the
Galactic ridge emission further, the summed spectrum of the quiescent
state in 1996/1997 is added to the simultaneous fit.

The final combined analysis of all 2005, 2007, and 2011 April data
(compare Table~\ref{tab:obsids}) results in a remarkable fit with an
$\chi_\mathrm{red}^2 \approx 1.10$ for 3651 dof. The
flux independent parameters, $N_\mathrm{H}$, $E_\mathrm{fold}$, and
$kT$, the outburst dependent parameters (iron line equivalent widths),
the Galactic ridge parameters, and the calibration constant are
presented in Table~\ref{tab:alldatafitpars}.

\begin{figure}
 \includegraphics[width=\columnwidth]{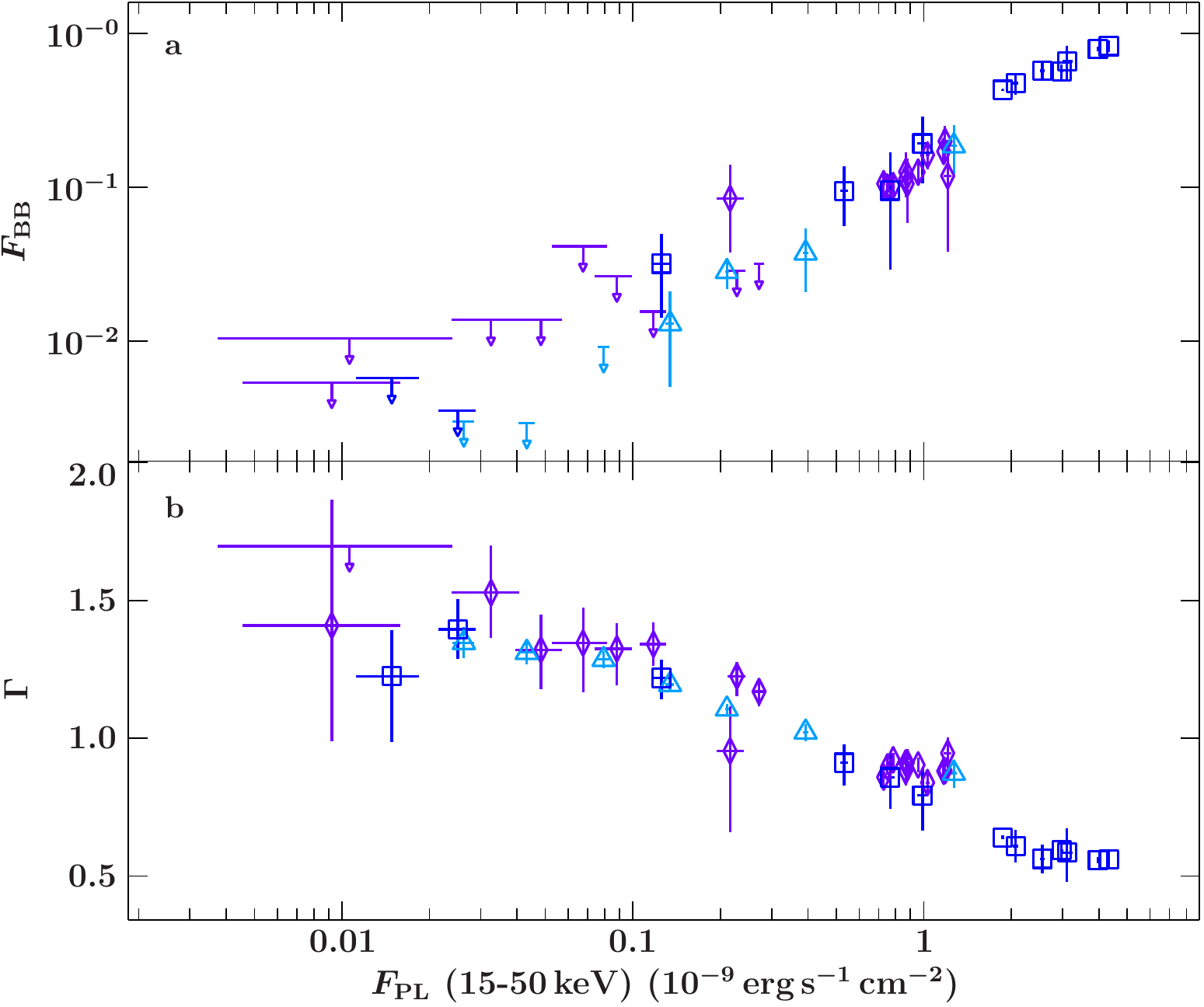}
 \caption{\textbf{a)} Black body flux, $F_\mathrm{BB}$ in
   $10^{-9}\,\mathrm{erg}\,\mathrm{s}^{-1}\,\mathrm{cm}^{-2}$, and
   \textbf{b)} photon index, $\Gamma$, both strongly correlate with the
   power law flux, $F_\mathrm{PL}$. Light blue triangles and dark blue
   squares represent data from the decays of the 2005 and 2007 outbursts,
   respectively, purple diamonds are from the rise of the 2011 April
   outburst (Table~\ref{tab:obsids} and Fig.~\ref{fig:asm0507}).}
 \label{fig:paramevol}
\end{figure}

Due to the combined fits of all available data, the uncertainties of
all parameters can be reduced significantly by comparing
Table~\ref{tab:alldatafitpars} with Table~\ref{tab:combinedparams}.
This is not only seen for the flux independent parameters, but also
for the continuum parameters allowed to vary for each observation,
especially for those at low source luminosity, where only four
parameters are required to describe the source spectrum. Apart from an
instrumental parameter, the PCA background factor, $b_\mathrm{PCA}$,
the three physical parameters describing the evolution of the spectral
continuum are the photon index $\Gamma$, the black body flux,
$F_\mathrm{BB}$, and the power law flux, $F_\mathrm{PL}$. As
shown in Fig.~\ref{fig:paramevol} the three parameters are strongly
correlated: the photon index decreases with $\log F_\mathrm{PL}$,
i.e., the source hardens with luminosity, and the black body
flux increases linearly with powerlaw flux. Note that the trend from
the rise of the 2011 outburst (purple diamonds) is equal to the trends
of the declines of both outbursts in 2005 and 2007. Thus, the spectral
shape does not depend on the sign of the time derivative of the flux.

The relationships shown in Fig.~\ref{fig:paramevol} can be used to
reduce the number of parameters needed to describe the spectrum of
\gro to one, $F_\mathrm{PL}$, if the flux independent parameters are
fixed to the values listed in Table~\ref{tab:alldatafitpars}. A fit to
the data shown in Fig.~\ref{fig:paramevol} gives the following
empirical relationships:
\begin{align}\label{eq:GammaFpl}
 \Gamma &= a + b
 \ln\left(\frac{F_\mathrm{PL}}{10^{-9}\,\mathrm{erg}\,\mathrm{s}^{-1}\,\mathrm{cm}^{-2}}\right)
   \\ 
\intertext{and}
 \label{eq:FbbFpl}
 \frac{F_\mathrm{BB}}{10^{-9}\,\mathrm{erg}\,\mathrm{s}^{-1}\,\mathrm{cm}^{-2}} &= c + d \frac{F_\mathrm{PL}}{10^{-9}\,\mathrm{erg}\,\mathrm{s}^{-1}\,\mathrm{cm}^{-2}}
\end{align}
with $a = 0.834(10)$, $b = -0.184(8)$ and $c = -0.019(7)$, $d =
0.191(9)$ ($\chi^2_\mathrm{red} = 1.82$, 38\,dof and
$\chi^2_\mathrm{red} = 1.09$, 24\,dof, respectively).
Using the $c$ parameter of the straight line describing the black body
flux one finds that the black body starts to contribute significantly
once the power law reached a 15--50\,keV flux of $10(4) \times
10^{-11}\,\mathrm{erg}\,\mathrm{s}^{-1}\,\mathrm{cm}^{-2}$. This
corresponds to a total luminosity of $13(5) \times
10^{35}\,\mathrm{erg}\,\mathrm{s}^{-1}$ assuming a distance of
5.8\,kpc \citep{riquelme2012a}.

In both fits, the data points resulting from the \Suzaku-spectra were
ignored, since they differ significantly from the best-fit to the
correlations. This difference is due to energy calibration differences
between \RXTE and \Suzaku, which lead to a slight change in the fitted
powerlaw photon index $\Gamma$ of $\Delta\Gamma\sim0.1$. Such
differences in photon index are typical between missions. See, e.g.,
\citealt{kirsch2005a}, who find differences up to $\Delta\Gamma=0.2$
between different missions in their analysis of the Crab pulsar and
nebula.

\begin{table}
   \caption{Source flux independent parameters, the iron line equivalent
     widths, the parameters of the Galactic ridge emission, and the
     HEXTE calibration constant as determined from the combined spectral
     analysis ($\chi^2_\mathrm{red}/\mathrm{dof} = 1.10/3651$).}
\renewcommand{\arraystretch}{1.3}

 \label{tab:alldatafitpars} 
 \begin{tabular}{llll}
   \hline\hline
   \texttt{TBnew} & $N_\text{H}$ & $1.547^{+0.019}_{-0.023}$ $\times 10^{22}$\,cm\,$^{-2}$ \\
\texttt{BBODY} & $kT$ & $1.833^{+0.015}_{-0.017}$ keV \\
\texttt{CUTOFFPL} & $E_\text{fold}$ & $15.92^{+0.24}_{-0.27}$ keV \\
iron line & $W_\text{2005}$ & $65^{+10}_{-10}$ eV \\
 & $W_\text{2007}$ & $27.9^{+2.4}_{-2.3}$ eV \\
 & $W_\text{2011}$ & $83^{+5}_{-5}$ eV \\
constants\tablefootmark{a} & $c_\text{HEXTE}$ & $0.859^{+0.009}_{-0.009}$  \\
 & $c_\text{XRT}$ & $0.806^{+0.009}_{-0.009}$  \\
 & $c_\text{XIS0}$ & $0.889^{+0.007}_{-0.007}$  \\
 & $c_\text{XIS1}$ & $0.936^{+0.007}_{-0.007}$  \\
 & $c_\text{XIS3}$ & $0.887^{+0.007}_{-0.007}$  \\
 & $c_\text{PIN}$ & $1.000^{+0.010}_{-0.010}$  \\
 & $c_\text{GSO}$ & $1.17^{+0.09}_{-0.09}$  \\
\texttt{BREMSS}\tablefootmark{b} & $kT$ & $3.4^{+0.5}_{-0.5}$ keV \\
 & $F_\text{3-10\,keV}$ & $4.25^{+0.20}_{-0.23}$ $\times 10^{-12}$\,erg\,s$^{-1}$\,cm$^{-2}$ \\
blended iron lines\tablefootmark{b} & $E$ & $6.349^{+0.026}_{-0.031}$ keV \\
 & $\sigma$ & $0.53^{+0.06}_{-0.05}$ keV \\
 & $F$ & $2.39^{+0.17}_{-0.15}$ $\times 10^{-4}$\,photons\,s$^{-1}$\,cm$^{-2}$ \\

   \hline
 \end{tabular}
\tablefoot{
  \tablefoottext{a}{detector calibration constants, $c$, are given
    relative to the \RXTE-PCA}
  \tablefoottext{b}{component belongs to the Galactic ridge emission
    and is applied to \RXTE-PCA only (Eq.~\ref{eq:modelgre})}
}
\end{table}

\subsection{Cyclotron Resonant Scattering Feature} \label{sec:spectral:crsf}

\begin{figure}
 \includegraphics[width=\columnwidth]{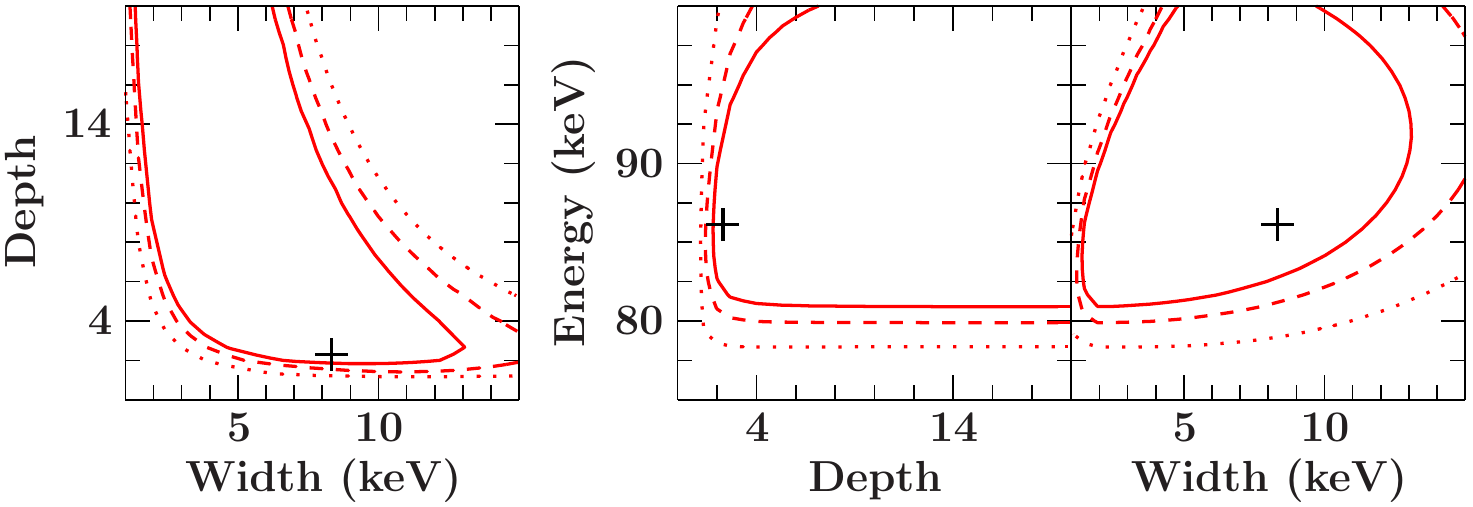}
  \caption{Contour maps between the \texttt{CYCLABS}-parameters used
    to model the absorption above 75\,keV. The solid line represents
    the $1\sigma$ contour, the dashed line $2\sigma$, and the dotted
    line $3\sigma$.} 
  \label{fig:confmapscyc}
\end{figure}

As discussed in Sect.~\ref{sec:spectral:cont}, a possible absorption-like
feature above 75\,keV is visible in the \Suzaku-GSO and \RXTE-HEXTE
spectra (Figs.~\ref{fig:specmodels} and ~\ref{fig:combinedfits},
respectively). Initial attempts to fit this
feature were unsuccessful due to its rather low significance
and strong correlations between the continuum parameters and
the line parameters. The flux dependency of the
continuum parameters (Eq.~\ref{eq:GammaFpl} and~\ref{eq:FbbFpl}) and
the large number of flux independent spectral components
(Table~\ref{tab:alldatafitpars}) significantly reduce the number of
continuum parameters. To describe the line, the continuum model of
Eq.~\eqref{eq:model} is modified with a multiplicative cyclotron line
model of the form
\begin{equation} \label{eq:cyclabs}
  \texttt{CYCLABS}(E) = \exp\left(-\frac{\tau (W E/E_\mathrm{cyc})^2}{(E-E_\mathrm{cyc})^2 + W^2}\right)
\end{equation}
where $E_\mathrm{cyc}$ is the centroid energy, $W$ the width, and
$\tau$ the optical depth of the cyclotron line. Performing a fit to
the simultaneous \Suzaku and \RXTE spectrum measured in 2007 results
in a satisfactory description of the cyclotron line. The best-fit
parameters together with the historic values determined using
\CGRO-OSSE data are shown in Table~\ref{tab:cyclo}. The depth is fixed
to the value found by \cite{shrader1999a} since the signal-to-noise
ratio in the line core is close to unity, allowing the modelled depth
to take any arbitrary value. Although the remaining parameters are
consistent with previous results, the data quality at these high
energies does, however, not allow constraining any parameters well. In
addition, the $\chi^2$-contours of the line parameters plotted in
Fig.~\ref{fig:confmapscyc} reveal strong correlations. The centroid
energy of the absorption feature is consistent with nearly any value
above 75\,keV. Even a combined fit of all \RXTE data with the
\texttt{CYCLABS} model added does not improve the fits nor does it constrain
the parameters better. Despite the large number of observations
presented here, it is therefore difficult to distinguish between a
cyclotron resonant scattering feature or a modification of the shape
of the high energy cutoff away from a pure exponential cutoff at these
energies.

\begin{table}
 \centering
 \caption{Parameters of the possible cyclotron resonant scattering
   feature around 88\,keV. See text for further discussion.} 
 \label{tab:cyclo}
 \begin{tabular}{llll}
   Parameter & This work & \cite{shrader1999a} & Unit \\
   \hline\hline
   $E_0$ & $86^{+7}_{-5}$ & $88^{+2.4}_{-2.4}$ & keV \\
   $W$ & $8^{+6}_{-4}$ & 20 (fixed) & keV \\
   $\tau$ & 2.3 (fixed) & $2.3^{+6}_{-6}$ & \\
   \hline
 \end{tabular}
\end{table}

\section{The 2011 December Outburst of \gro} \label{sec:dec2011}

In 2011 December, \gro underwent an even stronger outburst than the
outburst of 2007, which was analyzed  in the previous sections. The
peak flux in \Swift-BAT of $\sim$320\,mCrab was reached right
at the expected date around MJD\,55913 predicted by the orbital
ephemeris from Sect.~\ref{sec:timing}.

\Swift ToO observations obtained by the authors as a result of this
prediction are listed in Table~\ref{tab:obsids}. No further X-ray
data of sufficient quality of this outburst are available in the
archive and, thus, the high energy part of the spectrum above 10\,keV
is not available for analysis. The measurements made during this
outburst, which did not enter the analysis in the previous section,
provide a good test for the overall reliability of the simple
continuum model behavior found from the earlier outbursts. Fixing the
flux independent continuum parameters and $N_\mathrm{H}$-value to the
values listed in Table~\ref{tab:alldatafitpars}, and expressing
$\Gamma$ and $F_\mathrm{BB}$ as a function of power law flux through
Eq.~\ref{eq:GammaFpl} and~\ref{eq:FbbFpl}, the two free fit parameters
are the power law flux and the combined iron line equivalent width.

\begin{figure}
  \includegraphics[width=\columnwidth]{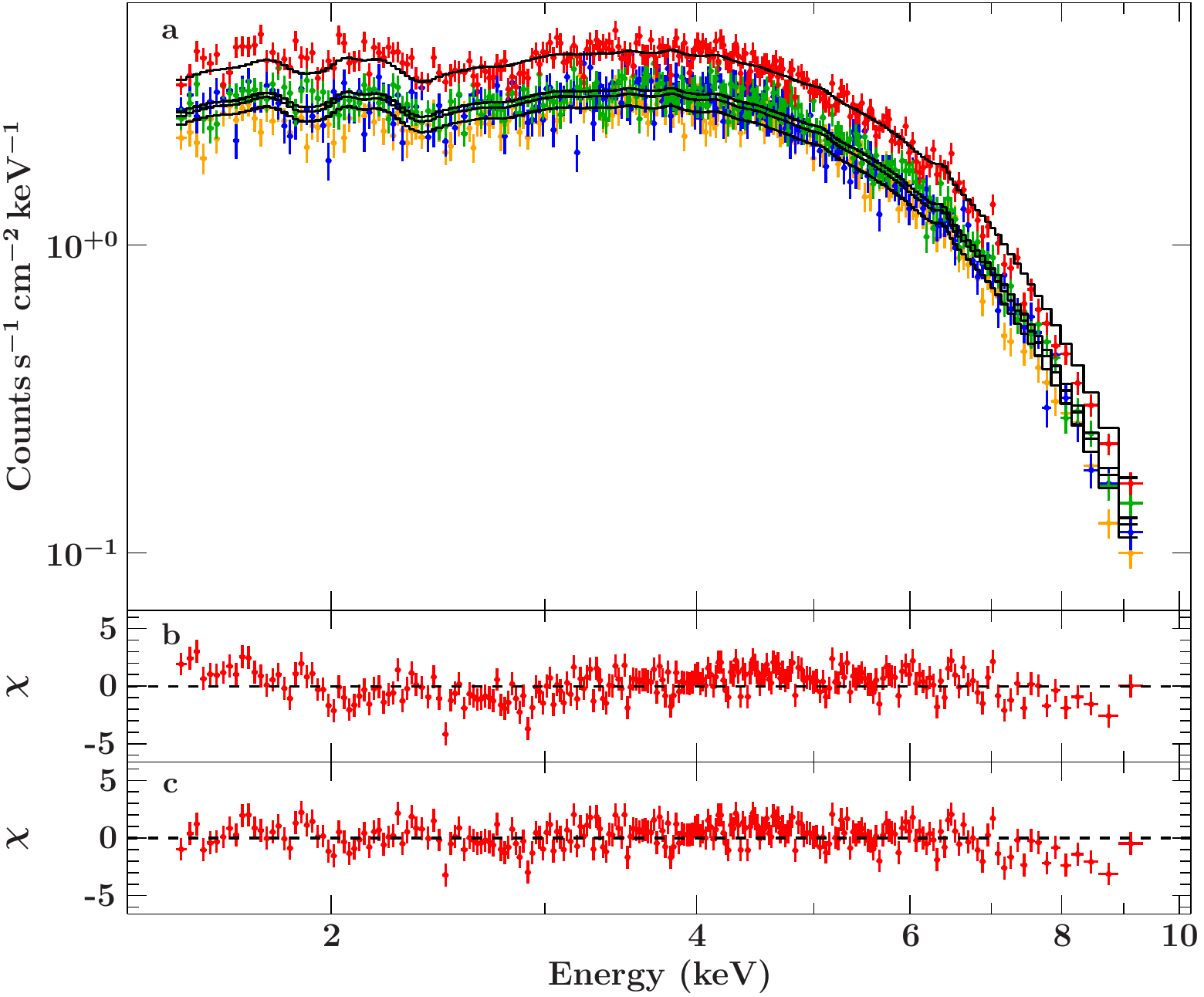}
  \caption{\textbf{a)} Best fit of the four \Swift-XRT spectra of
    the 2011 December outburst. \textbf{b)} Combined residuals
    after applying the same model to the data as used in the previous
    analysis (see text). The only three free spectral parameters are the
    power law flux, the absorption column density, and the Fe K$\alpha$
    equivalent width. \textbf{c)} Combined residuals after applying
    a partial coverer model (see text and Eq.~\ref{eq:particalcov}).}
  \label{fig:specdec2011}
\end{figure}

An initial combined fit of all four \Swift-XRT spectra yields an
unacceptable fit of $\chi^2_\mathrm{red} = 3.7$. This result is due to
residuals below 4\,keV, which is strong evidence for a change in the
absorption column density. Allowing this parameter to vary results in
a reasonable fit of $\chi^2_\mathrm{red} = 1.23$ for 934\,dof. There
are, however, slight deviations from a perfect fit left
(Fig.~\ref{fig:specdec2011}b). If a partial coverer model of the form
\begin{multline} \label{eq:particalcov}
F_\mathrm{ph, partial} = \left((1-f) \times
  \texttt{TBnew}(N_{\mathrm{H},1}) +\right. \\
\left. f \times \texttt{TBnew}(N_{\mathrm{H},2})\right) F_\mathrm{ph, model}(E)
\end{multline}
where $f$ is the covering factor, is applied to the model, the quality
of the fit is slightly increased ($\chi^2_\mathrm{red} = 1.12$ for
932\,dof). Note, however, that the spectra can also be described by
adding a secondary black body component with a temperature of
0.25\,keV. This latter approach results in an even better description
of the data ($\chi^2_\mathrm{red} = 1.07$ for 922\,dof). The
parameters of both possible models are shown in
Table~\ref{tab:parsdec2011} and the spectra with the secondary black
body model applied are shown in Fig.~\ref{fig:specdec2011}. Data from
the outbursts of \gro in 2012 show that there is indeed an additional
soft excess instead of a partial covering material (see
Sect.~\ref{sec:giant2012} and~\ref{sec:conclude} below).

Twelve days after the last \Swift pointing during the outburst, on
MJD~55932, \Swift started to observe \gro again. Until MJD~55958 six
additional observations were made (see Table~\ref{tab:obsids2}). Even
though the \Swift-BAT did not detect the source
(Fig.~\ref{fig:bat12}), \Swift-XRT images clearly reveal an X-ray
source at the position of \gro, indicating that the neutron star was
still accreting. Although the extracted \Swift-XRT spectra have a few
hundred photons only, by performing a combined fit of all six spectra
and using the same model as described above (without need of a partial
coverer model or secondary black body), the data can be well
described with an $\chi^2_\mathrm{red} = 1.05$ for 65\,dof. The final
parameters are listed in Table~\ref{tab:parspost2011}. The hydrogen
column density is slightly larger compared to the value
during the main outburst in 2011 December. This can be explained by
the position of the neutron star on the orbit, which is behind the
companion as seen from Earth. Here, the line of sight crosses a much
larger region within the system (see Fig.~\ref{fig:orbitcover}), and
thus increased absorption caused by the normal orbital modulation of
$N_\mathrm{H}$ is expected (see \citealt{hanke2010a} and
\citealt{hanke2011a} for a discussion of similar behavior in the HMXBs
\object{Cygnus~X-1} and \object{LMC~X-1}).

To conclude, the model of \gro found using the earlier data analyzed
in the previous sections works for other outbursts as well. \gro's
behavior is therefore independent of the outburst history.

\begin{table}
 \centering
 \caption{Spectral parameters of a fit to the \Swift-XRT spectra of
   the 2011 December outburst using a partial coverer model
   ($\chi^2_\mathrm{red} = 1.12$, 932\,dof) and a secondary black
   body component ($\chi^2_\mathrm{red} = 1.07$, 929\,dof).}
 \label{tab:parsdec2011}  
 \renewcommand{\arraystretch}{1.3}
 \begin{tabular}{lllll}
   component & & part. cov. & $2^\text{nd}$ black body & unit \\
   \hline\hline
   \texttt{TBnew} & $N_\text{H,1}$ & $1.45^{+0.20}_{-0.33}$ & $2.99^{+0.20}_{-0.19}$ & $10^{22}\,\text{cm}^{-2}$ \\
 & $N_\text{H,2}$ & $8.9^{+3.4}_{-2.7}$ & - & $10^{22}\,\text{cm}^{-2}$ \\
 & $f$ & $0.27^{+0.13}_{-0.07}$ & - &  \\
\texttt{BBODY\tablefootmark{a}} & $kT$ & - & $0.244^{+0.017}_{-0.019}$ & keV \\
 & $F_\text{BB,1}$ & - & $0.85^{+0.21}_{-0.16}$ & $10^{-9}$\,erg\,s$^{-1}$\,cm$^{-2}$ \\
 & $F_\text{BB,2}$ & - & $0.58^{+0.16}_{-0.14}$ & $10^{-9}$\,erg\,s$^{-1}$\,cm$^{-2}$ \\
 & $F_\text{BB,3}$ & - & $0.53^{+0.16}_{-0.14}$ & $10^{-9}$\,erg\,s$^{-1}$\,cm$^{-2}$ \\
 & $F_\text{BB,4}$ & - & $0.35^{+0.13}_{-0.11}$ & $10^{-9}$\,erg\,s$^{-1}$\,cm$^{-2}$ \\
\texttt{PL}\tablefootmark{a} & $F_\text{PL,1}$ & $4.65^{+0.10}_{-0.08}$ & $4.61^{+0.08}_{-0.08}$ & $10^{-9}$\,erg\,s$^{-1}$\,cm$^{-2}$ \\
 & $F_\text{PL,2}$ & $3.52^{+0.08}_{-0.08}$ & $3.54^{+0.06}_{-0.06}$ & $10^{-9}$\,erg\,s$^{-1}$\,cm$^{-2}$ \\
 & $F_\text{PL,3}$ & $3.33^{+0.08}_{-0.08}$ & $3.33^{+0.08}_{-0.08}$ & $10^{-9}$\,erg\,s$^{-1}$\,cm$^{-2}$ \\
 & $F_\text{PL,4}$ & $2.87^{+0.06}_{-0.06}$ & $2.93^{+0.06}_{-0.06}$ & $10^{-9}$\,erg\,s$^{-1}$\,cm$^{-2}$ \\
iron line & $W$ & $17^{+10}_{-10}$ & $16^{+10}_{-10}$ & eV \\

   \hline
 \end{tabular}
 \tablefoot{
   \tablefoottext{a}{$F_{\mathrm{PL},i}$ is the 15--50\,keV
     flux and $F_\mathrm{BB,i}$ is the bolometric black body flux of
     the $i$th \Swift observation of 2011 December
     (Table~\ref{tab:obsids}), scaled to the PCA via the
     detector calibration constant $c_\mathrm{XRT}$
     (Table~\ref{tab:combinedparams}).}
 }
\end{table}

\begin{table}
 \centering
 \caption{Spectral parameters of a fit to the \Swift-XRT spectra after
   the 2011 December outburst ($\chi^2_\mathrm{red} = 1.01$, 65\,dof).}
 \label{tab:parspost2011}  
 \renewcommand{\arraystretch}{1.3}
 \begin{tabular}{lllll}
   \hline\hline
   \texttt{TBnew} & $N_\text{H}$ & $4.3^{+0.7}_{-0.6}$ $\times 10^{22}\,\text{cm}^{-2}$ \\
\texttt{PL}\tablefootmark{a} & $F_\text{PL,1}$ & $5.9^{+1.8}_{-1.6}$ $\times 10^{-12}$\,erg\,s$^{-1}$\,cm$^{-2}$ \\
 & $F_\text{PL,2}$ & $7.8^{+1.9}_{-1.7}$ $\times 10^{-12}$\,erg\,s$^{-1}$\,cm$^{-2}$ \\
 & $F_\text{PL,3}$ & $4.0^{+1.4}_{-1.2}$ $\times 10^{-12}$\,erg\,s$^{-1}$\,cm$^{-2}$ \\
 & $F_\text{PL,4}$ & $2.9^{+0.9}_{-0.8}$ $\times 10^{-12}$\,erg\,s$^{-1}$\,cm$^{-2}$ \\
 & $F_\text{PL,5}$ & $0.28^{+0.26}_{-0.19}$ $\times 10^{-12}$\,erg\,s$^{-1}$\,cm$^{-2}$ \\
 & $F_\text{PL,6}$ & $1.8^{+0.8}_{-0.7}$ $\times 10^{-12}$\,erg\,s$^{-1}$\,cm$^{-2}$ \\

   \hline
 \end{tabular}
 \tablefoot{
   \tablefoottext{a}{$F_{\mathrm{PL},i}$ is the 15--50\,keV
     flux of the $i$th \Swift observation after 2011 December 
     (Table~\ref{tab:obsids}), scaled to the PCA via the
     detector calibration constant $c_\mathrm{XRT}$
     (Table~\ref{tab:combinedparams}).}
 }
\end{table}

\section{Predicting the Unpredictable: The 2012 November Type II
  Outburst of \gro} \label{sec:giant2012}

\begin{figure}
 \includegraphics[width=\columnwidth]{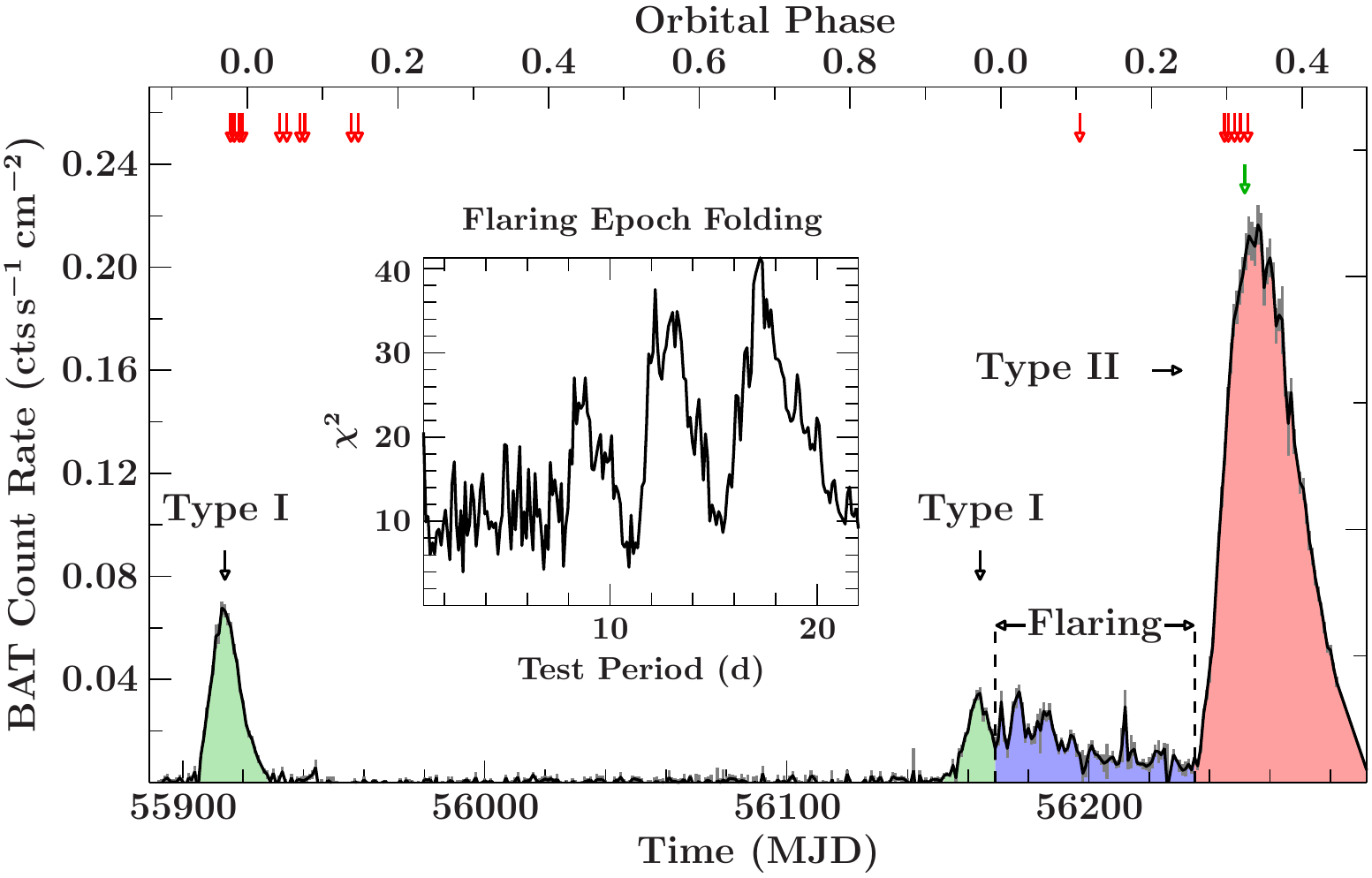}
  \caption{\Swift-BAT light curve showing the 2011 December and the
    2012 August outburst, which occurred as predicted by the orbital
    ephemeris (green region), and the unexpected flaring activity
    after the latter outburst (blue region) followed by a giant type
    II outburst at an orbital phase of nearly 0.3 (red region). 
    Times of observation by \Swift (red) and \Suzaku (green) are
    shown as arrows on top (compare Table~\ref{tab:obsids}). The
    inset shows the epoch folding result of the flaring part of the
    light curve.} 
  \label{fig:bat12}
\end{figure}

Although \gro seems to be predictable in both, the occurrence of
outbursts and the spectral shape, an unexpected and unpredictable
behavior occurred right after the outburst in 2012 August, which
itself occurred at expected time. Instead of fading into
quiescence the outburst decay lasted several weeks, during
which several flares were detected (see Fig.~\ref{fig:bat12}). An epoch
folding \citep{leahy1983a} of the flaring activity in the light curve
reveals significant variability on timescales on the order of 10 d and
above (see Fig.~\ref{fig:bat12}, inset).

\Swift observed \gro during the flaring activity on 2012 September
26/27. Due to the low flux
($\sim$$1.14\times10^{-9}\,\mathrm{erg}\,\mathrm{s}^{-1}\,\mathrm{cm}^{-2}$),
the XRT data were grouped to achieve a a minimum signal-to-noise ratio
of 5. The same model as used for the previous outbursts was then
applied to the data, which resulted in an good fit
($\chi^2_\mathrm{red} = 0.97$, 117\,dof). The corresponding spectral
parameters are listed in Table~\ref{tab:parssep2012}.
Although the best-fit iron line equivalent
width is much larger than during previous outburst, the lower
confidence limit is still in agreement with the value during the
2011~December outburst.

\begin{table}
 \centering
 \caption{Spectral parameters of a fit to the \Swift-XRT spectrum
   during the flaring activity after the 2012 August outburst
   ($\chi^2_\mathrm{red} = 0.97$, 117\,dof).}
 \label{tab:parssep2012} 
\renewcommand{\arraystretch}{1.3}
 \begin{tabular}{lllll}
   \hline\hline
   \texttt{TBnew} & $N_\text{H}$ & $2.04^{+0.23}_{-0.22}$ $\times 10^{22}\,\text{cm}^{-2}$ \\
\texttt{PL}\tablefootmark{a} & $F_\text{PL}$ & $1.81^{+0.11}_{-0.11}$ $\times 10^{-9}$\,erg\,s$^{-1}$\,cm$^{-2}$ \\
iron line & $W$ & $110^{+90}_{-90}$ eV \\

   \hline
 \end{tabular}
 \tablefoot{
   \tablefoottext{a}{ $F_\mathrm{PL}$ is the 15--50\,keV power law
     flux, scaled to the PCA via the detector calibration
     constant $c_\mathrm{XRT}$ (Table~\ref{tab:combinedparams}).
 }
}

\end{table}

After the flares, on 2012 November 13, the BAT instrument onboard
\Swift was triggered by a sudden flux increase of \gro
\citep{atel4573}. The flux reached 1\,Crab within a few days,
indicating a further outburst of the source. This flux is about three
times higher than the strongest outburst detected previously except of
the discovery outburst \citep{shrader1999a} and nearly one order of
magnitude higher than the mean maximal flux over all outbursts detected
by ASM and BAT (Fig.~\ref{fig:asm0507}). According to the orbital
parameters listed in Table~\ref{tab:orbit} the unexpected outburst
occured close to orbital phase 0.3 \citep[see also][]{atel4561}. This
unusual behavior is typical for type II outbursts, which have been seen
in many other Be X-ray binaries such as A0535+26
\citep{caballero2012a} or EXO~2030+375 \citep{klochkov2008b}.

\begin{figure}
 \includegraphics[width=\columnwidth]{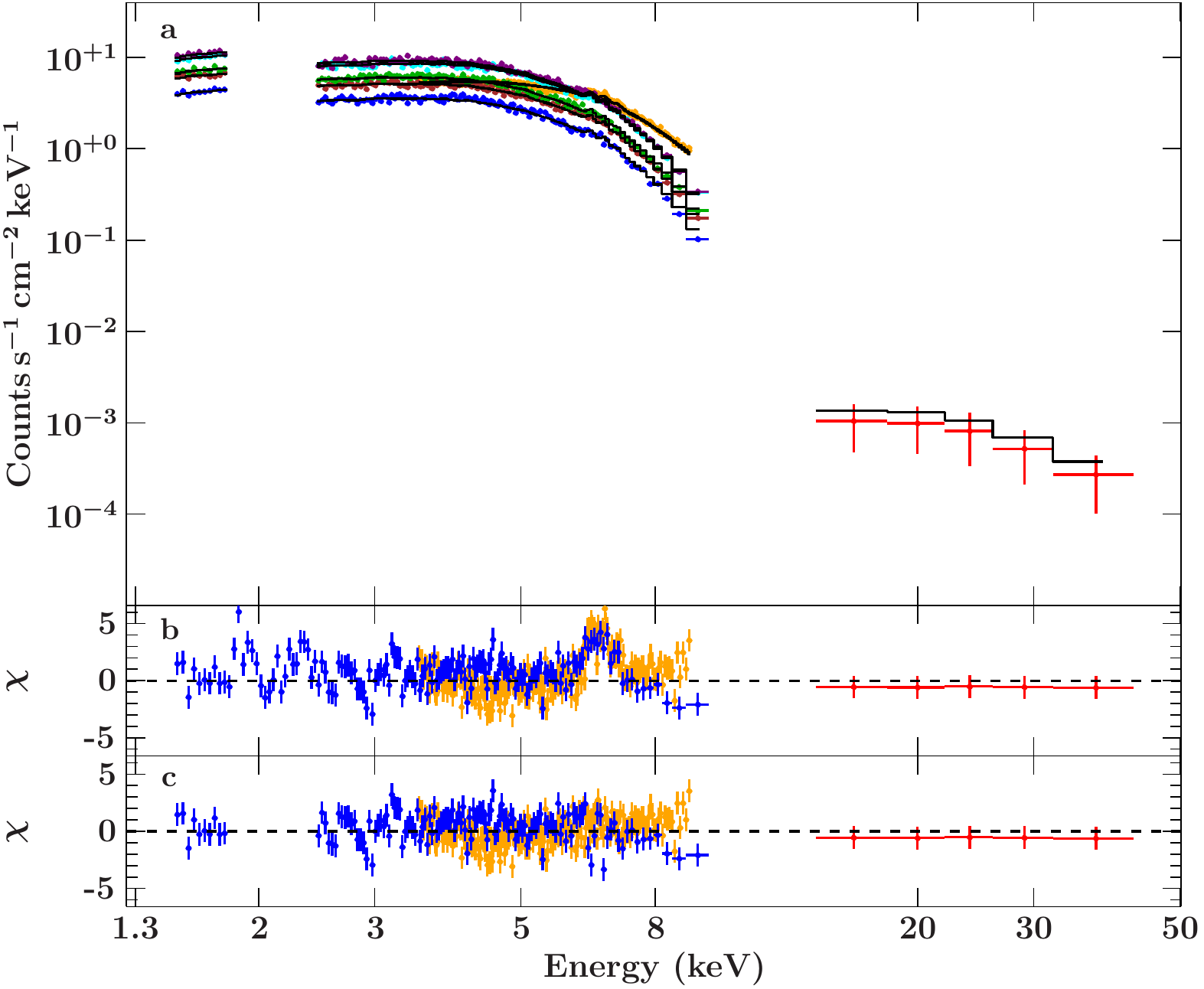}
 \caption{\textbf{a)} The five \Swift-XRT (blue), \Swift-BAT (red) and
   \Suzaku-XIS spectra
   of the type II 2012 November outburst of \gro. Spectral channels are
   rebinned for display purposes. \textbf{b)} Combined residuals after applying
   the same model to the data as used in the previous analysis including the
   secondary black body (see text).
   \textbf{c)} Best-fit combined residuals after excluding calibration
   features between 1.7 and 2.4\,keV and adding three narrow Fe lines at 6.4,
   6.67\,keV, and 7\,keV (compare Fig.~\ref{fig:linesnov2012}).}  \label{fig:specnov2012}
\end{figure}

The spectra of \gro as observed by \Swift-XRT, \Swift-BAT and \Suzaku-XIS
are shown in Fig.~\ref{fig:specnov2012}. The model applied to the data is
the same as used for the previous outbursts. The \Swift spectra are in
good agreement with this model after excluding calibration features at
$\sim$1.8\,keV and $\sim$2.2\,keV, accounting for the calibration of
the Si K edge and the Au M edge, respectively \citep{godet2007a,hurkett2008a}.
The BAT-spectrum was rebinned to a signal-to-noise ratio of 1.5 in the energy
range of 14 to 40\,keV. Due to the high flux level of the source
data from \Suzaku-XIS were extracted avoiding regions with more than 2\%
pile-up fraction.

An initial fit reveals residuals, which are most prominent in \Suzaku-XIS and
is of similar shape to the residuals seen in the 2011 December outburst
(compare Fig.~\ref{fig:specdec2011}). A partial covering model is not able to
fit the residual structure. Using a secondary black body component results in
a better description
of the data. Its temperature of around 0.39\,keV is similar to the value
found in the 2011 December data.
A complex structure below 3.5\,keV remains in the residuals of \Suzaku-XIS,
which is
not present in \Swift-XRT data. Since the overall continuum shape is modelled
successfully, XIS-data below 3.5\,keV were removed from the analysis for the
time being.

\begin{figure}
 \includegraphics[width=\columnwidth]{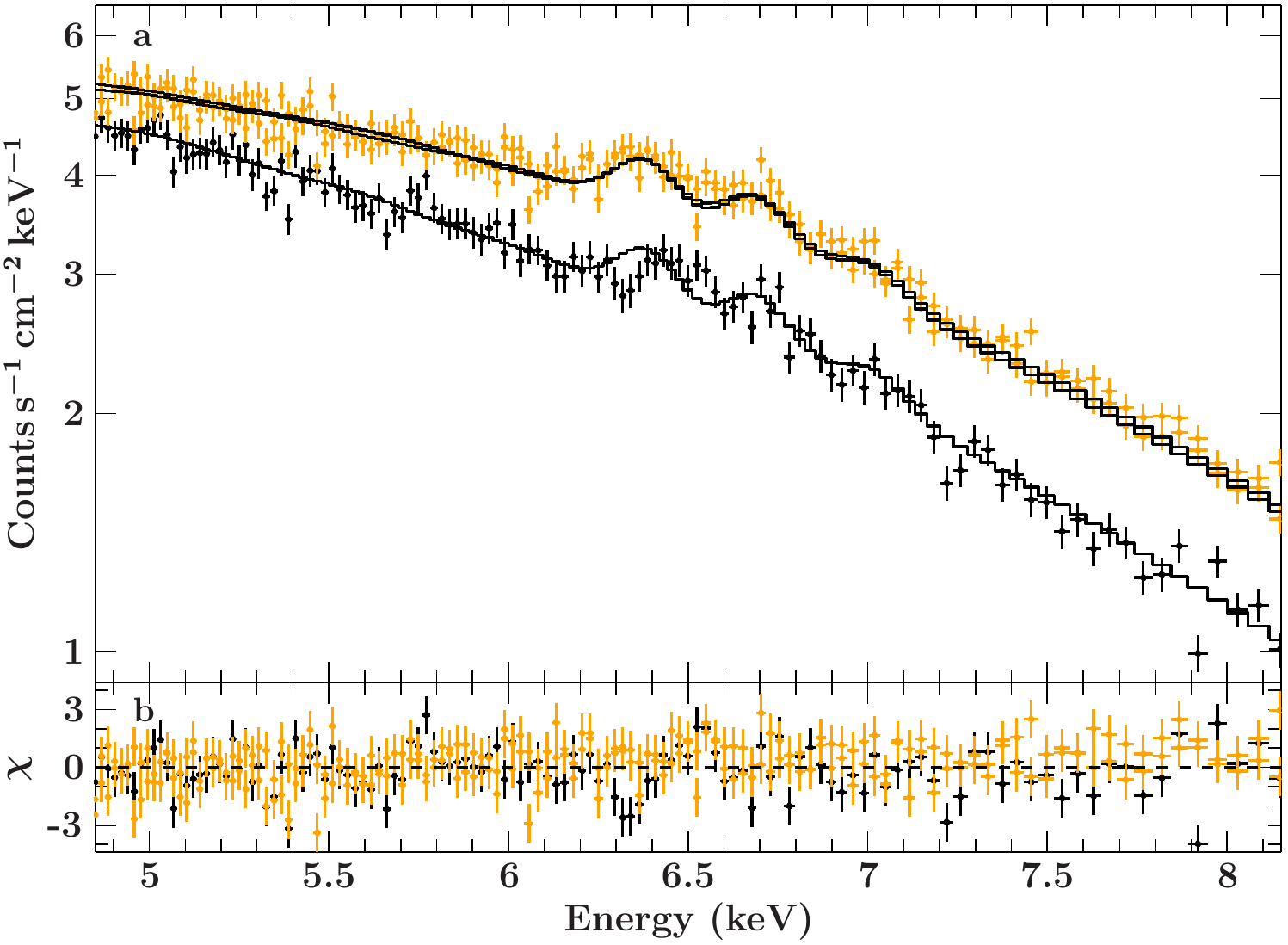}
 \caption{\textbf{a)} The iron line complex in \Suzaku-XIS (orange: XIS0/3,
   black: XIS1). The neutral, hydrogen- and helium-like emission lines
   are clearly visible as well as shift in the line centroid energies between
   XIS1 and XIS0/3. \textbf{b)} Residuals of the fit (see text).}
   \label{fig:linesnov2012}
\end{figure}

In addition to the continuum model, an unresolved residual
structure consistent with the presence of three narrow Fe emission
lines at 6.4, 6.67 and 7\,keV is visible in the spectra
(see Fig.~\ref{fig:linesnov2012}), corresponding
to neutral Fe K$\alpha$ fluorescence and recombination lines from
H-like and He-like iron. The centroid energies of the fluorescent lines, which
have been fixed relative to neutral iron emission at 6.4\,keV, are shifted by
$\sim+0.1$\,keV between the front illuminated XIS1 and the back illuminated
XIS0 and XIS3. This is probably due to calibration issues, such as a gain shift.
To ensure that the line parameters are not affected by this energy shift,
data from XIS1 have been excluded in the final fit.
With these modifications, a good description of the data is obtained
($\chi^2_\mathrm{red} = 1.08$, 2582\,dof). The equivalent widths of these
lines together with the fluxes, the secondary black body parameters and
absorption column density are listed in Table~\ref{tab:parsnov2012}.

During the 2012 November outburst the hydrogen column density is
comparable to the preceding flaring activity and the 2007 December outburst,
which itself is in agreement with the galactic hydrogen column density
(Tables~\ref{tab:parssep2012} and~\ref{tab:combinedparams}). As the
observations during and after the 2011 December outburst show, the
column density might still change between outbursts
(Tables~\ref{tab:parsdec2011} and~\ref{tab:parspost2011}). The extreme flux
increase during the 2012 giant outburst requires a large amount of
material. This is in contrast to the relatively low hydrogen column density.
The ionized iron in the system indicates, however, the presence of
significant amounts of almost fully ionized plasma, probably due to
photoionization caused by the very luminous neutron star, which would not
be picked up in absorption by X-ray spectroscopy. This would effectively
cause a decrease of the measured neutral hydrogen column density down to the
galactic value.

Performing a pulse period analysis of the five \Swift-XRT lightcurves
during the giant outburst, which have been corrected for binary motion
using the parameters listed in Table~\ref{tab:orbit}, reveals a
significant faster rotation period of the neutron star of
$P=93.6483(7)\,\mathrm{s}$ and a spin-up of
$\dot{P} = -0.60(4) \times 10^{-7}\,\mathrm{s}\,\mathrm{s}^{-1}$.

\begin{table}
 \centering
 \caption{Spectral parameters of a fit to the five \Swift-XRT, \Swift-BAT, and
   \Suzaku-XIS spectra of the giant 2012 November outburst
   ($\chi^2_\mathrm{red} = 1.08$, 2582 dof).} 
 \label{tab:parsnov2012} 
\renewcommand{\arraystretch}{1.3}
 \begin{tabular}{lllll}
   \hline\hline
   \texttt{TBnew} & $N_\text{H}$ & $1.86^{+0.24}_{-0.25}$ $\times 10^{22}\,\text{cm}^{-2}$ \\
\texttt{PL}\tablefootmark{a} & $F_\text{PL,1}$ & $9.82^{+0.11}_{-0.11}$ $\times 10^{-9}$\,erg\,s$^{-1}$\,cm$^{-2}$ \\
 & $F_\text{PL,2}$ & $10.46^{+0.13}_{-0.11}$ $\times 10^{-9}$\,erg\,s$^{-1}$\,cm$^{-2}$ \\
 & $F_\text{PL,3}$ & $11.55^{+0.13}_{-0.13}$ $\times 10^{-9}$\,erg\,s$^{-1}$\,cm$^{-2}$ \\
 & $F_\text{PL,4}$ & $14.69^{+0.14}_{-0.14}$ $\times 10^{-9}$\,erg\,s$^{-1}$\,cm$^{-2}$ \\
 & $F_\text{PL,5}$ & $14.31^{+0.16}_{-0.16}$ $\times 10^{-9}$\,erg\,s$^{-1}$\,cm$^{-2}$ \\
 & $F_\text{PL,Suz.}$ & $11.71^{+0.08}_{-0.08}$ $\times 10^{-9}$\,erg\,s$^{-1}$\,cm$^{-2}$ \\
\texttt{BBODY}\tablefootmark{a} & $kT$ & $0.390^{+0.017}_{-0.014}$ keV \\
 & $F_\text{BB,1}$ & $0.541^{+0.042}_{-0.042}$ $\times 10^{-9}$\,erg\,s$^{-1}$\,cm$^{-2}$ \\
 & $F_\text{BB,2}$ & $0.591^{+0.042}_{-0.042}$ $\times 10^{-9}$\,erg\,s$^{-1}$\,cm$^{-2}$ \\
 & $F_\text{BB,3}$ & $0.620^{+0.045}_{-0.045}$ $\times 10^{-9}$\,erg\,s$^{-1}$\,cm$^{-2}$ \\
 & $F_\text{BB,4}$ & $0.779^{+0.045}_{-0.045}$ $\times 10^{-9}$\,erg\,s$^{-1}$\,cm$^{-2}$ \\
 & $F_\text{BB,5}$ & $0.88^{+0.06}_{-0.06}$ $\times 10^{-9}$\,erg\,s$^{-1}$\,cm$^{-2}$ \\
 & $F_\text{BB,Suz.}$ & $0.64^{+0.32}_{-0.32}$ $\times 10^{-9}$\,erg\,s$^{-1}$\,cm$^{-2}$ \\
iron line & $W_\text{neutral}$ & $30^{+4}_{-4}$ eV \\
 & $W_\text{H-like}$ & $32^{+5}_{-5}$ eV \\
 & $W_\text{He-like}$ & $17^{+5}_{-5}$ eV \\
constants & $c_\mathrm{XIS0}$ & $0.787^{+0.006}_{-0.006}$  \\

   \hline
 \end{tabular}
 \tablefoot{
   \tablefoottext{a}{$F_{\mathrm{PL},i}$ is the 15--50\,keV
     flux and $F_\mathrm{BB,i}$ is the bolometric flux of the secondary
     black body of the $i$th \Swift observation of 2012 November as listed
     in Table~\ref{tab:obsids}, scaled to the PCA via the
     detector calibration constant $c_\mathrm{XRT}$ and $c_\mathrm{XIS3}$
     (Table~\ref{tab:combinedparams}). The calibration constant 
     $c_\mathrm{XIS0}$ had to be refitted. $F_{\mathrm{PL,Suz.}}$ and
     $F_{\mathrm{BB,Suz.}}$ are the corresponding fluxes obtained by the
     \Suzaku observation.}
 }
\end{table}

\section{Conclusions} \label{sec:conclude}

\begin{figure}
 \includegraphics[width=\columnwidth]{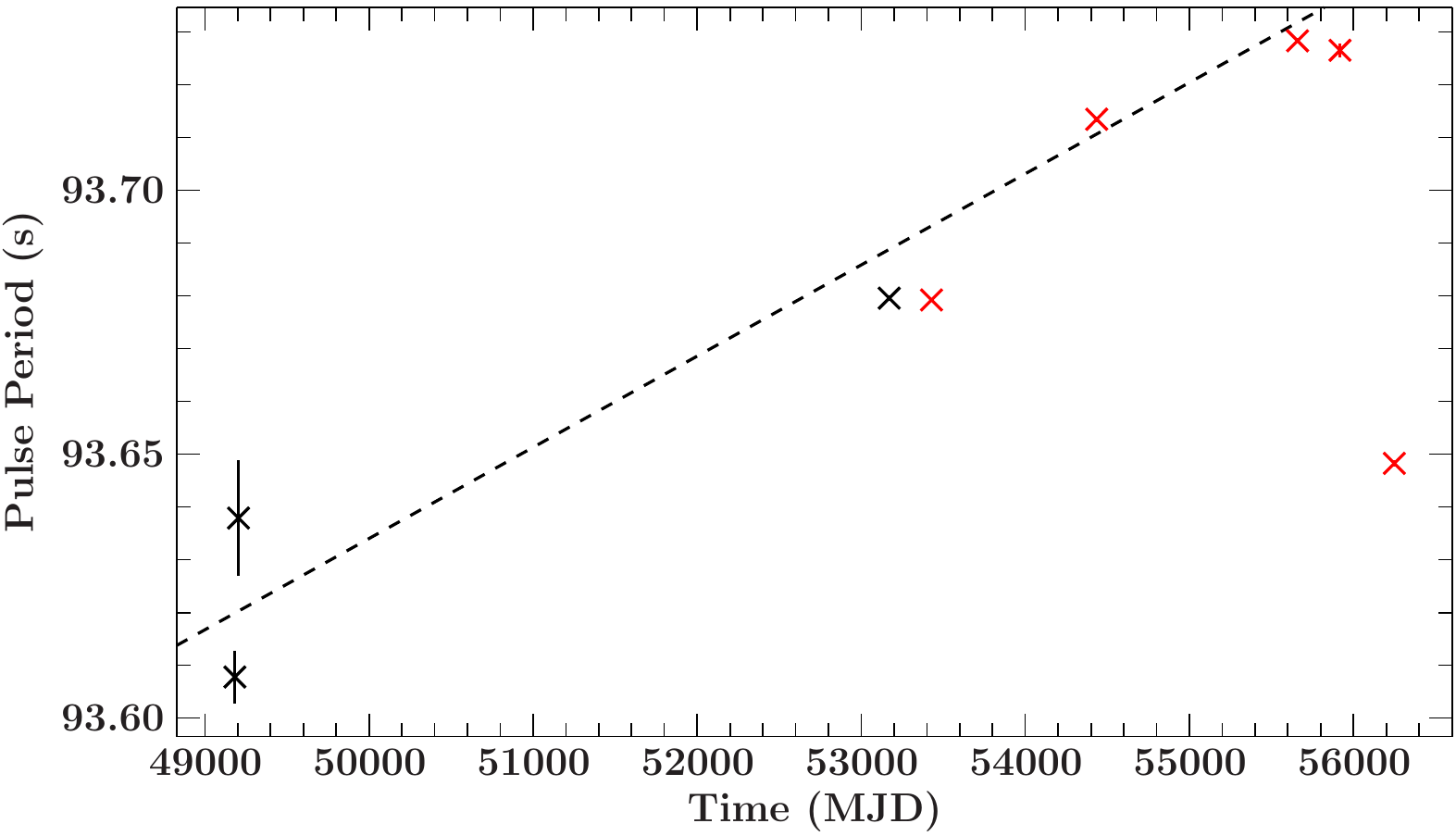}
 \caption{Spin period evolution of \gro since its discovery. Black:
   historic data taken from \citet{stollberg1993a},
   \citet{shrader1999a}, and \citet{coe2007a}. Red: results of this
   work. The latest period measurement was taken during the ``giant''
   2012 outburst. To guide the eye, the dashed line shows a
   spin-down of $2\times10^{-10}\,\mathrm{s}\,\mathrm{s}^{-1}$.}
  \label{fig:spinevol}
\end{figure}

In the previous sections, all available \RXTE observations of \gro
including three outbursts in 2005, 2007, and 2011 April as well as an
observation campaign during source quiescence in 1996/1997 were
analyzed. In addition to these data \Swift and \Suzaku observations
during the 2007 outburst were combined with the \RXTE analysis, and
\Swift observations of the 2011 December and ``giant'' 2012 November
outbursts were compared to the spectral results of the previous
outbursts.

By performing a detailed pulse arrival times analysis the orbital
period of the binary and the time of periastron passage were improved
(see Table~\ref{tab:orbit}). In addition, a slight non-linear spin-up
of the neutron star in the order of
$10^{-14}\,\mathrm{s}\,\mathrm{s}^{-2}$ was detected during high
luminosity of the 2007 outburst. This pulse ephemeris implies a spin-up
rate of $-2 \times 10^{-8}\,\mathrm{s}\,\mathrm{s}^{-1}$ at maximum
luminosity around MJD~54428. During the giant 2012 outburst, the
measured value of $-6 \times 10^{-8}\,\mathrm{s}\,\mathrm{s}^{-1}$ is
three times higher compared to the 2007 value. This difference is
consistent with theory, where the spin-up rate $\dot{P}$ is connected
to the luminosity $L$ via \citep{ghosh1979b}
\begin{equation}
 -\dot{P} \propto L^\alpha
\end{equation}
with $\alpha=1$ for wind- and $\alpha=6/7$ for disk-accretion. The
maximum luminosity during the giant 2012 outburst was about three times
larger than during the 2007 outburst.

The evolution of the period over the decades as shown in
Fig.~\ref{fig:spinevol} reveals, however, that the
longterm evolution is dominated by a spin-down in the order of
$2\times10^{-10}\,\mathrm{s}\,\mathrm{s}^{-1}$. A similar long term
behavior was detected in \object{A0535+26} by \citet[][see also
\citealt{bildsten1997a}]{camero-arranz2012a}. Such long term spin-down
might be evidence for the ``propeller'' regime, where matter near the
neutron star gets expelled, removing angular momentum and causing the
neutron star to spin-down
\citep{illarionov1975a,frank1992a,bildsten1997a}. In the propeller
regime, the magnitude of spin down is
\begin{equation}
 \dot{P} = \frac{10 \pi \mu^2}{G M^2 R^2}
\quad
\text{where}
\quad
\mu=\frac{1}{2} BR^3,
\end{equation}
where $B$ is the magnetic field strength on the polar caps of the
neutron star and where all other symbols have their usual meaning.
Assuming a mass $M=1.4\,M_\odot$ and a radius $R = 10\,\mathrm{km}$ of
the neutron star and $\dot{P}$ as estimated from
Fig.~\ref{fig:spinevol}, the surface magnetic field strength of \gro
is $B\sim2.6\times10^{12}\,\mathrm{G}$. This is a reasonable value for
an accreting neutron star in a HMXB and implies a fundamental CRSF at
30\,keV in the X-ray spectrum. Note, however, that since the spin-up
phases during outbursts lead to a lower observed long-term spin-down,
the derived magnetic field strength without taking this spin-up into
account is most likely a lower limit.

Using the precise orbital solution allows a detailed study of the
connection of outbursts with the orbit. In case of \gro all outbursts
detected in \RXTE-ASM and \Swift-BAT until 2012 May are well connected
to the periastron passage, with the peak flux occurring close to
periastron (Fig.~\ref{fig:outburst_phase}). This behavior is in line
with results from a recent study of Be outburst behavior by
\citet{okazaki2012a}. These authors show that in Be systems type~I
outbursts occur close to periastron, where the neutron star can
accrete from a tidally truncated Be-disk. According to
\citet{okazaki2012a}, type~I outbursts occur regularly only in systems
with high eccentricity ($\gtrsim$0.6), where the Be disk is truncated
close to the critical Roche Lobe radius. The regularity of type~I
outbursts is increased in systems with large orbital periods, where
the disturbance of the Be-disk is minimized. Both conditions are
fulfilled in \gro, for which \citet{okazaki2001a} calculated the disk
of the donor to be truncated at the 7:1 or 8:1 resonance and concluded
that this system should show regular type~I outbursts. A further
prediction is that type~II ``giant'' outbursts can happen if there is
a misalignment between the Be-disk and the orbital plane
\citep{okazaki2012a}. These outbursts would then be triggered, e.g.,
by increased activity of the donor star as seen, for example, in
A0535+26 \citep[e.g.,][]{yan2012a}. There are indications that such a
misalignment is also present in \gro, since the type~I outburst
maximum happens slightly before periastron
(Fig.~\ref{fig:outburst_phase}). Note that another Be source,
\object{2S1845$-$024}, shows similar behavior \citep{finger1999a},
although here the regular type~I outbursts occur slightly after
periastron.

Between the ordinary type~I outburst in August 2012 and the giant
type~II outburst several flares were detected in \Swift-BAT (see
Fig.~\ref{fig:bat12}). An epoch folding \citep{leahy1983a} of that time range revealed
oscillations in the order of $\sim 10$\,d (see inset of
Fig.~\ref{fig:bat12}). The origin of these oscillations is unknown.

As shown in Sect.~\ref{sec:spectral:cont}, the spectrum of \gro can be
well described by a cut-off power law with an additional black body
component, with some spectral parameters being independent of flux and
outburst (Table~\ref{fig:combinedfits}). The physical reason might be
found in the unique properties of the neutron star in \gro, such as
its magnetic field strength, mass and radius. Without a working
physical model describing the spectra of accreting neutrons stars it
is, however, not possible to investigate this aspect any further.
Nevertheless, \gro is an ideal candidate to test future physical
models, which are still in development \citep[see][and references
therein]{becker2007a}. For a distance of 5.8\,kpc and a neutron star
with 10\,km radius, the maximum observed source flux during the
2007 December outburst corresponds to an emission area of around 1.5\%
of the neutron star's surface.
Even during the giant 2012 November outburst the derived area fraction
was only 5\%.
These values are in good agreement with estimates of the hot spot size
at the magnetic poles \citep[e.g.,][]{gnedin1973a,ostriker1973a}.

Similar correlations as the one discovered between the photon index
$\Gamma$ and the black body flux $F_\mathrm{BB}$ with the power law
flux $F_\mathrm{PL}$ (Fig.~\ref{fig:paramevol}) have been seen in
other Be X-ray binaries as well \citep[see, e.g.,][for a recent
discussion]{reig2013a}. These authors analyzed the PCA-data of
the 2007 outburst of \gro and found a similar correlation between
the powerlaw and the source luminosity. In addition, they discovered
a correlation between the folding energy and the photon index.
Taking the HEXTE-data into account, Fig.~\ref{fig:confmaps} shows,
however, that the data is consistent with a constant folding energy.

In case of \gro, the spectrum hardens with increasing luminosity.
Recent theoretical work by \citet{becker2012a}
describes several spectral states depending on the source luminosity,
where sources with luminosities below the Eddington limit are expected
to show this kind of correlation. Assuming a source distance of
5.8\,kpc \citep{riquelme2012a} and using the maximum flux values given
in Table~\ref{tab:combinedparams}, the peak luminosity of \gro is around
$3\times10^{37}\,\mathrm{erg}\,\mathrm{s}^{-1}$. This value is
close to the critical luminosity reported by \cite{becker2012a}, which
scales like $L_\mathrm{crit} \sim 1.5 \times 10^{37} B^{16/15}_{12}$.
Thus, \gro is likely to be a subcritical accretor during type~I
outbursts.

The most remarkable result of the analysis of
the giant type~II outburst of \gro is that the spectral model found for
type~I outbursts (Eq.~\ref{eq:model} with flux independent
parameters from Table~\ref{tab:combinedparams} and flux correlations
from Eqs.~\ref{eq:GammaFpl} and~\ref{eq:FbbFpl}) is also able to
describe the type~II hard X-ray ($>$10\,keV) spectrum. The main
contribution in that energy range is the powerlaw component, which
emerges from inverse Compton effects in the accretion column far above
the neutron star's surface. To describe the soft X-ray spectrum, however,
an additional soft component is needed (the second black body). This
suggests a change in the accretion mechanisms near the neutron star's
surface, where Comptonization processes take place and shocks may form
as described in \citet{becker2012a}. Following the findings of these
authors, the derived luminosity of \gro during the giant outburst is
close to $10^{38}\,\mathrm{erg}\,\mathrm{s}^{-1}$, which indicates that the
neutron star is accreting supercritically. Thus, the spectral change in
the soft X-rays might be due supercritical accretion. A similar black
body component had to be introduced for the 2011 December
outburst (see Table~\ref{tab:parsdec2011}). Compared to the additional
soft component of the 2012 giant outburst, its bolometric flux
contributes much less. That indicates that the source was close to or
in the transition from the sub- to the supercritial accretion regime
during the 2011 December outburst. These findings are in agreement with
and confirm the state changes in neutron star Be X-ray binaries as
proposed by \cite{reig2013a}, where they found a softening of the
spectrum close or above the critical luminosity.

Neutral material in the vicinity of the neutron star is
believed to be the origin of observed fluorescence lines, e.g., the
6.4\,keV iron K$\mathrm{\alpha}$ line. For $N_\mathrm{H}\lesssim
10^{23}\,\mathrm{cm}^{-2}$, self-absorption is not important
and the iron line equivalent width and the hydrogen column
density scale roughly linearly \citep{eikmann2012a}. The generally low
observed equivalent widths indicate that the fluorescence line originates
in the absorbing material. In most observations, $N_\mathrm{H}$ is in
agreement with the interstellar value, although some intrinsic
absorption cannot be ruled out. During the 2012 August outburst, the
best fit Fe K$\alpha$ equivalent width is much larger (although with
large uncertainties). This larger equivalent width could be due to the
addition of a line from reflection from, e.g., an enhanced Be disk.
If the X-rays are reflected by material, the
equivalent width increases significantly up to a factor of 100.

The data analyzed within this paper can not confirm the claimed
cyclotron line at 88\,keV by \citet[][see
Sect.~\ref{sec:spectral:crsf}]{shrader1999a}, who argued that
this feature is rather the second harmonic. Although there are
slight deviations from the continuum visible in both, \RXTE-HEXTE and
\Suzaku-GSO data, the signal does not allow to constrain any
parameters. During the ``giant'' 2012 November outburst of \gro, the
signal quality of the \Suzaku-PIN and -GSO data could be used to
further investigate this claimed high energy cyclotron line. These
data were not used here since no background data were available at the
time of writing. However, \citet{atel4759} were able to fit the
spectra with preliminary background corrections. These authors claimed a
cyclotron feature between 74 and 78\,keV, close to the feature at 88\,keV
reported by \citet{shrader1999a}.

In conclusion, the fact that the spectrum of several outbursts of \gro
can be modelled with the single simple model of Eqs.~\eqref{eq:model},
\eqref{eq:GammaFpl}, and \eqref{eq:FbbFpl} shows that the shape and
behavior of the spectrum is well understood. Due to the flux
independent parameters and the flux correlations found during the
analysis (Table~\ref{tab:alldatafitpars} and Fig.~\ref{fig:paramevol})
\gro's spectrum at any time in the subcritial state can be described
given only the power law flux. This is a remarkable result which has
not been seen in any other
transient X-ray binary before. \gro therefore shows a well
predictable behavior in both, outburst dates and spectral shape.
Therefore this source is an ideal target to clarify more detailed
aspects of Be X-ray transients: what drives the peak flux of
outbursts? Are there correlations or changes in iron line flux and
absorption column density? How can physical models of accretion
explain the spectral shape of \gro? Are there other sources where
similar predictable behavior exists, and are there differences to
\gro? Does the spectral model of \gro work outside of the flux range
covered here?

\begin{acknowledgement}
  We thank the \RXTE-, \Swift- and \Suzaku-teams for their role in
  scheduling all observations used within this paper and for accepting
  our proposals. We especially thank Evan Smith for his help in
  scheduling the \RXTE observations in 2011 April. Many thanks to Hans
  Krimm for providing the BAT-spectrum during the 2012 giant outburst.
  We acknowledge
  funding by the Bundesministerium f\"ur Wirtschaft und Technologie
  under Deutsches Zentrum f\"ur Luft- und Raumfahrt grants 50OR0808,
  50OR0905, 50OR1113, and the Deutscher Akademischer Austauschdienst.
  This work has been partially supported by the Spanish Ministerio de
  Ciencia e Innovaci\'on through projects AYA2010-15431 and
  AIB2010DE-00054. All figures shown in this paper were produced using
  the \texttt{SLXfig} module, developed by John E. Davis. We thank the
  referee for her/his helpful comments and suggestions.
\end{acknowledgement}

\bibliography{mnemonic,aa_abbrv,./references}
\bibliographystyle{aa}

\end{document}